\documentclass[fleqn,usenatbib]{mnras}

\usepackage{newtxtext,newtxmath}

\usepackage[T1]{fontenc}
\usepackage{ae,aecompl}

\usepackage{graphicx}	
\usepackage{amsmath}	
\usepackage{amssymb}	
\usepackage{multicol}        
\usepackage{bm}		
\usepackage{pdflscape}	
\usepackage{soul}
\usepackage{enumitem}
\usepackage{pdflscape}	
\usepackage[dvipsnames]{xcolor}
\usepackage[normalem]{ulem}

\newcommand{\Mpc}{\mathrm{Mpc}}
\newcommand{\Myr}{\mathrm{Myr}}
\newcommand{\Gyr}{\mathrm{Gyr}}
\newcommand{\kpc}{\mathrm{kpc}}
\newcommand{\pc}{\mathrm{pc}}
\newcommand{\kmpers}{\mathrm{km} \, \mathrm{s}^{-1}}

\newcommand{\cmsquare}{\mathrm{cm}^{-2}}

\newcommand{\Msol}{\textup{M}_\mathrm{\sun}}

\newcommand{\K}{\mathrm{K}}
\newcommand{\xMsol}[2]{\ensuremath{{#1}\times 10^{#2} \,\Msol}}
\newcommand{\xScientific}[2]{\ensuremath{{#1} \times 10^{#2}}}

\newcommand{\Msolyr}{\textup{M}_\mathrm{\sun} \, \text{yr}^{-1}}

\newcommand{\mdm}{m_{\mathrm{DM}}}

\newcommand{\Mvir}{M_{200}}

\newcommand{\Mstar}{M_{\star}}

\newcommand{\rhalflight}{r_{1/2, \text{3D}}}
\newcommand{\vcirc}{v_{\text{circ}}}
\newcommand{\vphi}{v_{\phi, g}}
\newcommand{\vdisp}{\sigma_{\text{turb, g}}}
\newcommand{\vdispeff}{\sigma_{\text{eff}}}
\newcommand{\vR}{v_{R, g}}
\newcommand{\vrot}{v_{\text{rot}, g}}
\newcommand{\vrothi}{v_{\text{rot, \hi}}}

\newcommand{\xhi}{x_{\mathrm{\hi}}}
\newcommand{\Mhi}{M_{\mathrm{\hi}}}

\newcommand{\NhiaboveXX}[2]{\Nhi \geq \xScientific{#1}{#2}\, \cmsquare}

\newcommand{\Nhi}{N_{\mathrm{\hi}}}
\newcommand{\hinospace}{H\,\textsc{i}}
\newcommand{\hi}{H\,\textsc{i} }

\newcommand{\MstarMhi}{\Mstar - \Mhi}

\graphicspath{{./}{figs/}}

\title[The smallest \hi rotation curves]
{EDGE -- Dark matter or astrophysics? Breaking dark matter heating degeneracies with \hi rotation in faint dwarf galaxies}

\author[M. P. Rey et al.]
{Martin P. Rey,$^{1}$\thanks{E-mail: \href{martin.rey@physics.ox.ac.uk}{martin.rey@physics.ox.ac.uk}} Matthew D. A. Orkney,$^{2}$ Justin I. Read,$^{3}$ Payel Das,$^3$ Oscar Agertz$^4$, Andrew Pontzen$^5$, 
\newauthor Anastasia A. Ponomareva$^{1}$, Stacy Y. Kim$^{3}$ and William McClymont $^{3,6,7}$ 
\vspace{0.8mm}
\\
$^{1}$ Sub-department of Astrophysics, University of Oxford, DWB, Keble Road, Oxford OX1 3RH, UK \\ 
$^{2}$ Institut de Ciencies del Cosmos (ICCUB), Universitat de Barcelona (IEEC-UB), Martí i Franquès 1, E08028 Barcelona, Spain \\
$^{3}$ Department of Physics, University of Surrey, Guildford GU2 7XH, UK \\
$^{4}$ Lund Observatory, Division of Astrophysics, Department of Physics, Lund University, Box 43, SE-221 00 Lund, Sweden \\
$^{5}$ Department of Physics and Astronomy, University College London, London WC1E 6BT, UK \\
$^{6}$ Kavli Institute for Cosmology, University of Cambridge, Madingley Road, Cambridge, CB3 0HA, UK \\
$^{7}$ Cavendish Laboratory, University of Cambridge, Madingley Road, Cambridge, CB3 9BB, UK
}

\date{Submitted to MNRAS}

\pubyear{2024}

\begin{document}
\label{firstpage}
\pagerange{\pageref{firstpage}--\pageref{lastpage}}
\maketitle

\begin{abstract}  
Low-mass dwarf galaxies are expected to reside within dark matter haloes that have a pristine, `cuspy' density profile within their stellar half-light radii. This is because they form too few stars to significantly drive dark matter heating through supernova-driven outflows. Here, we study such simulated faint systems ($10^4 \leq \Mstar \leq 2 \times 10^6 \, \Msol$) drawn from high-resolution (3 pc) cosmological simulations from the `Engineering Dwarf Galaxies at the Edge of galaxy formation' (EDGE) project. We confirm that these objects have steep and rising inner dark matter density profiles at $z=0$, little affected by galaxy formation effects. But five dwarf galaxies from the suite also showcase a detectable \hi reservoir ($\Mhi \approx 10^{5}-10^{6} \, \Msol$), analogous to the observed population of faint, \hinospace-bearing dwarf galaxies. These reservoirs exhibit episodes of ordered rotation, opening windows for rotation curve analysis. Within actively star-forming dwarfs, stellar feedback easily disrupts the tenuous \hi discs ($\vphi \approx 10\, \kmpers$), making rotation short-lived ($\ll 150 \, \Myr$) and more challenging to interpret for dark matter inferences. In contrast, we highlight a long-lived ($\geq 500 \, \Myr$) and easy-to-interpret \hi rotation curve extending to $\approx 2\, \rhalflight$ in a quiescent dwarf, that has not formed new stars since $z=4$. This stable gas disc is supported by an oblate dark matter halo shape that drives high-angular momentum gas flows. Our results strongly motivate further searches for \hi in rotation curves in the observed population of \hinospace-bearing low-mass dwarfs, that provide a key regime to disentangle the respective roles of dark matter microphysics and galaxy formation effects in driving dark matter heating.
\end{abstract}




\section{Introduction} \label{sec:intro}
The existence of a significant amount of dark matter in our Universe is firmly established, with its gravitational influence leaving distinct signatures on the cosmic microwave background (e.g. \citealt{PlanckCollaboration2020}), the large-scale distribution of galaxies (e.g. \citealt{Alam2021}), and the dynamics of baryonic tracers in galaxies and galaxy clusters (e.g. \citealt{Zwicky1933, Rubin1980, Clowe2006}). But the microphysical nature of dark matter and its direct detection remains elusive, despite extensive efforts in the last decade (see e.g. \citealt{Schumann2019} for a review). This calls for a wide and thorough scan of parameter space to robustly remove alternatives, motivating complementary efforts across disciplines  (\citealt{Bertone2018}).

The latest data from galaxy counts (e.g. \citealt{Nadler2021MWConstraints}), stellar stream gaps (e.g. \citealt{Banik2021DMConstraints}), strong lensing (e.g. \citealt{Gilman2020, Hsueh2020}), the Ly$\,\alpha$ forest (e.g. \citealt{Irsic2017,Armengaud2017, Rogers2021}) or their combination (e.g. \citealt{Nadler2021Lensing+MW, Enzi2021}) all point to dark matter being a cold (i.e. non-relativistic at the time of decoupling) collisionless, particle. However, this stills leaves plenty of possible options for the physical nature of the constituent sourcing the dark matter gravitational field (e.g. supersymmetric weakly interacting massive particles, sterile neutrinos, axions, etc; see \citealt{Bertone2005, Bertone2018} for reviews). 

Galactic rotation curves are one of the first historical probes of dark matter and continue to play a key role in the effort to narrow down the available parameter space of models (e.g. \citealt{Rubin1970, Rubin1980, vanAlbada1985, deBlok2002, Oh2011, Oh2015, Lelli2016, Posti2019, ManceraPina2020}). In particular, rotation curves and \hi kinematics of small dwarf galaxies are particularly powerful. They can be used to determine the dark matter halo masses hosting small dwarf galaxies, directly constraining the low-mass end of the galaxy-halo connection and dark matter models that suppress small-scale power in the cosmological power spectrum (e.g. a warm or wave dark matter; \citealt{Polisensky2011, Anderhalden2013, Kennedy2014, Read2017, Nadler2021MWConstraints, Yasin2023, Sardone2023}). Rotation curves are also sensitive to the structure of the inner gravitational potential and can be used to infer the dark matter distributions and density profiles in dwarf galaxies (e.g. \citealt{Flores1994, Moore1994, deBlok2002, Oh2011, Oh2015, Ott2012, Iorio2017}). This in turn provides constraints on mechanisms heating dark matter, either dynamically or through microphysical particle interactions (e.g. annihilation or self-interactions).

Both of these features have historically garnered significant interest from the community because, at face value, they are discrepant with predictions assuming pure cold dark matter (CDM; see \citealt{Pontzen2014, Bullock2017, Sales2022} for reviews). CDM-only structure formation predicts many more bound dark matter subhaloes than observed satellite galaxies around the Milky Way and other nearby spiral galaxies -- the `missing satellite problem' \citep[e.g.][]{Moore1999,Klypin1999}. These same simulations predict centrally divergent dark matter density profiles inside dwarf galaxies, `cusps', whereas observations favour lower density `cores' -- the `cusp-core problem' \citep[e.g.][]{Flores1994, Moore1994, deBlok2002, Oh2011, Oh2015, Iorio2017}. Both problems can be solved by moving beyond the CDM assumption. For example, the cusp-core problem can be mitigated by making dark matter self-interacting \citep[e.g.][]{Burkert2000, Spergel2000} or fuzzy \citep[e.g.][]{Schive2014a, Veltmaat2018, Nori2021}, while satellite numbers can be reduced by suppressing small-scale cosmological power (e.g. \citealt{Boehm2014, Vogelsberger2019}).

However, these discrepancies can also be explained by a careful modelling of the physics of galaxy formation. In the case of the missing satellite problem, the solution involves a mix of accounting for observational completeness (e.g. \citealt{Kim2018}), star formation becoming inefficient in low mass haloes (e.g. \citealt{Efstathiou1992, Somerville2002, Sawala2016, Read2019SFRMatching}) and the tidal destruction of satellites on plunging orbits (e.g. \citealt{Read2006Tides,Garrison-Kimmel2017DiscDepletion}). In the case of the cusp-core problem, it can be solved by dark matter being dynamically heated during galaxy formation via repeated gas inflows and outflows \citep[e.g.][]{Navarro1996, Read2005, Pontzen2012}, or via dynamical perturbations induced by massive clumps or companions \citep[e.g.][]{El-Zant2001, Romano-Diaz2009, Goerdt2010, Nipoti2015, Orkney2021}. There is now compelling observational evidence that `dark matter heating' occurred in nearby dwarfs \citep[e.g.][]{Read2019DMHeating,Bouche2022, DeLeo2023}.

This makes testing dark matter models with rotation curves more ambiguous, as the effects of alternative dark matter models become degenerate with the physics of galaxy formation which remains challenging to model from first principles (see \citealt{Somerville2015, Naab2017} for reviews). This motivates us to find `clean' regimes to test models -- galaxies in which the rotation curve data are straightforward to interpret, and where dark matter and astrophysical models make testable predictions with minimal overlap. The best candidates for this are the smallest dwarf galaxies, where low stellar masses leave little opportunity for star formation and galaxy formation effects to impact the inner dark matter density profile \citep[e.g.][]{Teyssier2013, DiCintio2014, Chan2015, Tollet2016, Read2019DMHeating, Lazar2020, Orkney2021}. But these `ultra-faint' dwarfs are typically devoid of gas (\citealt{Geha2012, Putman2021}) and thus unsuitable for rotation curve analysis.

Excitingly, a growing number of isolated, gas-rich faint dwarfs have recently been reported, showcasing small but detectable \hi reservoirs ($\Mhi \approx 10^{5}-10^{6} \, \Msol$) that matches a faint stellar component ($10^4 \leq \Mstar / {\rm M}_\odot \leq 10^6$; \citealt{Irwin2007, Cole2014}; \citealt{McQuinn2015, McQuinn2020, McQuinn2021}; \citealt{Sand2015, Adams2018, Brunker2019, Janesh2019, Hargis2020, Bennet2022, Rhode2023}). Such low stellar masses (and thus low galaxy formation effects) and the presence of \hi (and thus of a dynamical tracer) could provide precisely the rotation curves needed to cleanly separate dark matter models from galaxy formation effects. Further, these objects are typically isolated `field' dwarfs, removing the need to model environmental effects from more massive hosts. 

However, a key puzzle remains before we can leverage this population of dwarf galaxies as a dark matter probe: none of them so far shows evidence for clear, ordered rotation in their \hi gas that can be easily exploited for dynamical modelling (\citealt{Bernstein-Cooper2014, Adams2018, McQuinn2021}). It remains unknown whether this is due to unfortunate inclination in the few examples observed (i.e. near face-on orientations), to the observational challenges associated with working with such small galaxies (e.g. \citealt{Read2016DwarfRCs, Verbeke2017,Oman2019, Downing2023} for discussions), or to an intrinsic lack of ordered rotation in the \hi gas altogether at this mass scale. 

In this paper, we address this puzzle using a suite of high-resolution cosmological `zoomed' simulations of faint dwarf galaxies from the EDGE project (introduced in \citealt{Agertz2020EDGE}). Our sample of simulated galaxies matches the observed population of isolated faint \hinospace-rich dwarfs, in both stellar masses, \hi masses and star formation activity (or lack thereof; \citealt{Rey2019UFDScatter, Rey2020, Rey2022EDGEHI}; Section~\ref{sec:methods}). In Section~\ref{sec:kinematics}, we extract their gas and \hi kinematics, highlighting multiple examples of rotationally supported gas kinematics and \hi rotation curves. This includes short-lived \hi discs rapidly dispersed by the energy input from massive stars, but also a long-lived example with near-circular rotation that could easily be modelled by standard mass-modelling tools (Section~\ref{sec:physicaldrivers}). We discuss the physical drivers of our results and their significance for future observational campaigns in Section~\ref{sec:conclusion}.

\section{The key regime of faint and \hi-rich dwarfs} \label{sec:methods}

We use the suite of faint ($10^4 \leq \Mstar \leq \xMsol{2}{6}$) simulated dwarf galaxies presented in \citet{Rey2022EDGEHI}, specifically focusing on the subset of five \hinospace-bearing dwarfs ($10^5 \leq \Mhi \leq 10^6 \, \Msol$). Next, we briefly summarize how each simulated galaxy is evolved to $z=0$ using cosmological, zoomed hydrodynamical simulations (see \citealt{Agertz2020EDGE, Rey2020} for more in-depth descriptions) and the characteristics of the simulated suite (see also \citealt{Rey2019UFDScatter, Rey2020, Rey2022EDGEHI, Orkney2021, Orkney2023}).

All galaxies are evolved to $z=0$ using cosmological zoomed, hydrodynamical simulations with the adaptive mesh refinement \textsc{ramses} code (\citealt{Teyssier2002}). The mass resolution inside the galaxy's Lagrangian region is enhanced using the \textsc{genetIC} software (\citealt{Stopyra2021}) to reach $\mdm = 960 \, \Msol$, while the hydrodynamical refinement strategy ensures a spatial resolution of 3 pc across the galaxy's interstellar medium (ISM; \citealt{Agertz2020EDGE}). The cosmological streaming of the Lagrangian patch of each galaxy is zeroed to reduce advection errors (\citealt{Pontzen2021}). We follow the formation of stars and the injection of energy, momentum, mass, and metals from asymptotic giant branch (AGB) stars, Type-II and Type-Ia supernovae (SNII, SNIa) according to \citet{Agertz2020EDGE}. We track the cooling of primordial and metal-enriched gas using equilibrium thermochemistry (\citealt{Courty2004}), accounting for on-the-fly self-shielding (\citealt{Aubert2010, Rosdahl2012}) and heating from a spatially uniform, time-dependent UVB (updated from \citealt{Haardt1996}; see \citealt{Rey2020} for further details). To derive \hi distributions, we evaluate the code's internal cooling function at every spatial position of the simulation and compute the neutral hydrogen fraction (\citealt{Rey2022EDGEHI}). We track dark matter haloes over time using the \textsc{hop} halo finder (\citealt{Eisenstein1998}) and construct merger trees using the \textsc{pynbody} and \textsc{tangos} libraries (\citealt{Pontzen2013, Pontzen2018}). We centre on our galaxies using the shrinking sphere algorithm (\citealt{Power2003}) on the dark matter and shift the velocity frame to put the central 1 kpc at rest. We  interpolate a single stellar population model (\citealt{Girardi2010}) over a grid of ages and metallicities to obtain the luminosities of individual stellar particles and compute the 3D stellar half-light radius, $\rhalflight$.

The \textsc{EDGE} simulated suite consists of ten low-mass dwarf galaxies ($\Mstar \leq \xMsol{2}{6}$) hosted in dark matter haloes with $10^9 \leq \Mvir \leq \xMsol{3}{9}$ at $z=0$ (\citealt{Rey2022EDGEHI}). At this mass-scale, all of our galaxies see their star formation truncated at high redshift ($z\geq 4$) following cosmic reionization, as their potential wells are then too shallow to accrete gas from the intergalactic medium (see e.g. \citealt{Efstathiou1992, Gnedin2000, Hoeft2006, Okamoto2008, Noh2014} for further discussions). 

Five of our dwarf galaxies assemble little dynamical mass at late times (i.e. after reionization) and have vanishing gas and \hi contents at $z=0$. Conversely, five others grow enough at late times to start re-accreting gas from the hot intergalactic medium and eventually host a detectable \hi reservoir at $z=0$ (\citealt{Rey2020,Rey2022EDGEHI}; see also \citealt{Ricotti2009, Benitez-Llambay2015, Fitts2017, Jeon2017, Ledinauskas2018, Benitez-Llambay2020, Benitez-Llambay2021, PereiraWilson2023} for further discussion of this re-accretion mechanism). 

The five \hinospace-bearing objects are the focus of this study. With $10^4 \leq \Mstar \leq \xMsol{2}{6}$ and $10^5 \leq \Mhi \leq 10^6 \, \Msol$, they provide excellent simulated analogues to the observed population of low-mass, \hinospace-bearing dwarfs in the $\MstarMhi$ plane (\citealt{Rey2022EDGEHI}, fig.~2). 

Furthermore, three out of five galaxies re-accreted their gas reservoirs early enough to re-ignite star-formation several billion years ago. Since their re-ignition, these galaxies have been forming stars with star formation rates (SFRs) $\approx 10^{-5} \, \Msolyr$ averaged over several billion years (\citealt{Rey2020}, fig.1), with instantaneous peaks up to $\text{SFR} \approx 10^{-4} \, \Msolyr$ (\citealt{Rey2020}, fig.7). These SFRs are similar to those measured in observed star-forming, low-mass dwarfs. For example, resolved colour-magnitude diagrams infer SFRs of a few $10^{-5} \, \Msolyr$ averaged over several billion years in Leo T (\citealt{Clementini2012, Weisz2012}), Leoncino (\citealt{McQuinn2021}), and Antlia B (\citealt{Hargis2020}). The H$\,\alpha$ detection of the single H~\textsc{ii} region of Leo P similarly implies SFR$\approx\xScientific{4}{-5} \, \Msolyr$ (\citealt{Rhode2013, McQuinn2015}), while the presence of blue helium-branch stars highlight recent, short peaks of $\approx 10^{-4} \, \Msolyr$ in Coma P (\citealt{Brunker2019}). Overall, our simulated galaxies are an excellent match to observed SFRs, \hi contents, stellar masses and stellar metallicities (\citealt{Collins2022}, fig.5) of faint dwarfs, making them an ideal platform to predict \hi kinematics in this regime. 

In addition to the three regularly star-forming dwarfs, another galaxy of our sample re-ignited star formation just before $z=0$ (at $z=0.03$, 500 Myr ago), after several billion years of quiescent but gas-rich evolution. The last galaxy is yet to reignite star formation despite hosting a significant gas reservoir (see \citealt{Rey2020, Benitez-Llambay2021, PereiraWilson2023} for the physical mechanisms affecting the timing of star formation reignition). 

These differences in star-formation activity are directly reflected in our simulated dwarfs' \hi properties. Star-forming dwarfs show strongly time varying, asymmetric \hi reservoirs that are often spatially offset from their stellar body (\citealt{Rey2022EDGEHI}). Quiescent dwarfs, by contrast, show more stable, more spherical and more aligned \hi contents over time (\citealt{Rey2022EDGEHI}). As we will see next, these distinctions are also reflected in the stability and structure of their gas and \hi kinematics.

\begin{figure}
  \centering
    \includegraphics[width=\columnwidth]{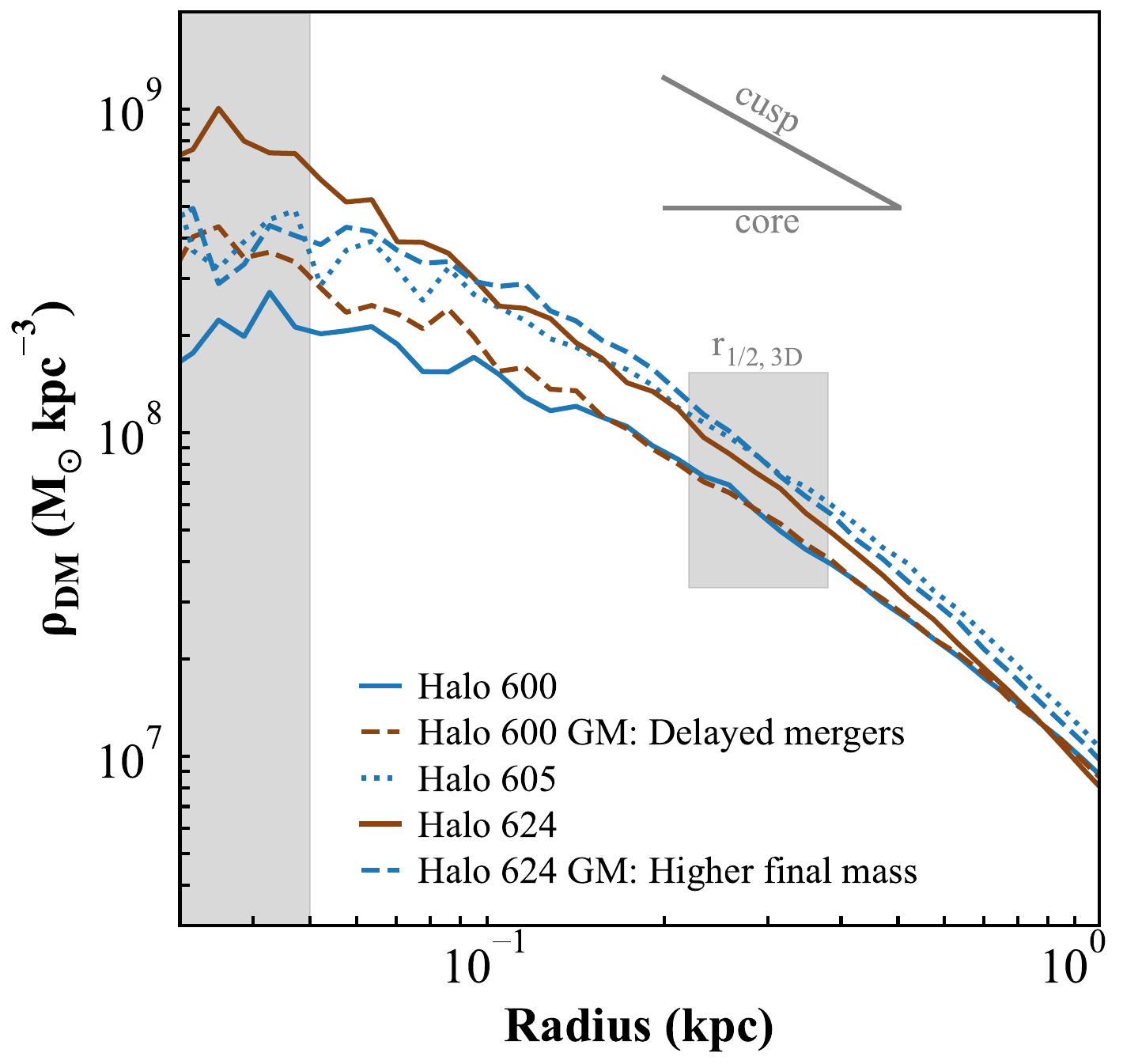}

    \caption{Dark matter density profiles across our suite of simulated, \hinospace-bearing faint dwarf galaxies. At this galactic mass scale, dynamical effects and supernova-driven outflows naturally arising in $\Lambda$CDM cosmologies can reduce central dark matter densities but are inefficient at forming large ($\rhalflight$-sized; grey box) and flat dark matter cores (indicative profile slopes marked in grey). Inferring the structures of dark matter haloes hosting these faint dwarfs, for example through \hi rotation (Figure~\ref{fig:hivelmaps}), thus holds great promises to distinguishing whether galaxy formation effects or new dark matter interactions drive dark matter heating in dwarfs. Interpreting the flattening of profiles at small radii (within marked grey box) is compromised by the limited resolution of the simulation.}
    \label{fig:dmdensity_profiles}
\end{figure}

Furthermore, all of our simulated dwarf galaxies exhibit steep and rising dark matter density profiles around $\rhalflight$. This is first shown in \citet{Orkney2021} using higher-resolution re-simulations ($\mdm = 120 \, \Msol$) of a subset of the galaxies studied here. This conclusion also holds at the resolution of this work ($\mdm = 960 \, \Msol$), with Figure~\ref{fig:dmdensity_profiles} showing the spherically averaged dark matter density profiles at $z=0$ using 100 log-spaced bins. All profiles are consistent with an increasing dark matter density towards the centre, with a `cuspy' logarithmic slope ($\approx-1$) until the limited resolution of the simulations starts affecting the dynamics of dark matter particles (grey band; see \citealt{Orkney2021}, appendix B for further discussion). 

Figure~\ref{fig:dmdensity_profiles} emphasizes that we have reached a critical regime. At this galactic mass scale, dynamical effects and star formation-driven outflows can lower central dark matter densities in the very centre of dwarfs, but are inefficient at forming large (i.e. $\approx \rhalflight$) and flat (i.e. constant-density) dark matter cores (see \citealt{Orkney2021} for further discussion). Any observational evidence for $\rhalflight$-sized dark matter cores in such faint dwarfs (e.g. \citealt{Amorisco2017Eri2, Sanders2018, Malhan2022CoresfromStreams}) thus becomes increasingly difficult to explain through purely astrophysical effects and could rather point to new dark matter physics (see e.g. \citealt{Orkney2022} for further discussion). The gas reservoirs of faint and \hinospace-bearing dwarfs provide a unique opportunity to obtain such observational insights provided that \hi kinematics can be harnessed to infer the structure of host dark matter haloes which we now quantify.

\section{Diverse and variable \hi kinematics in faint dwarfs} \label{sec:kinematics}

\begin{figure*}
  \centering
    \includegraphics[width=\textwidth]{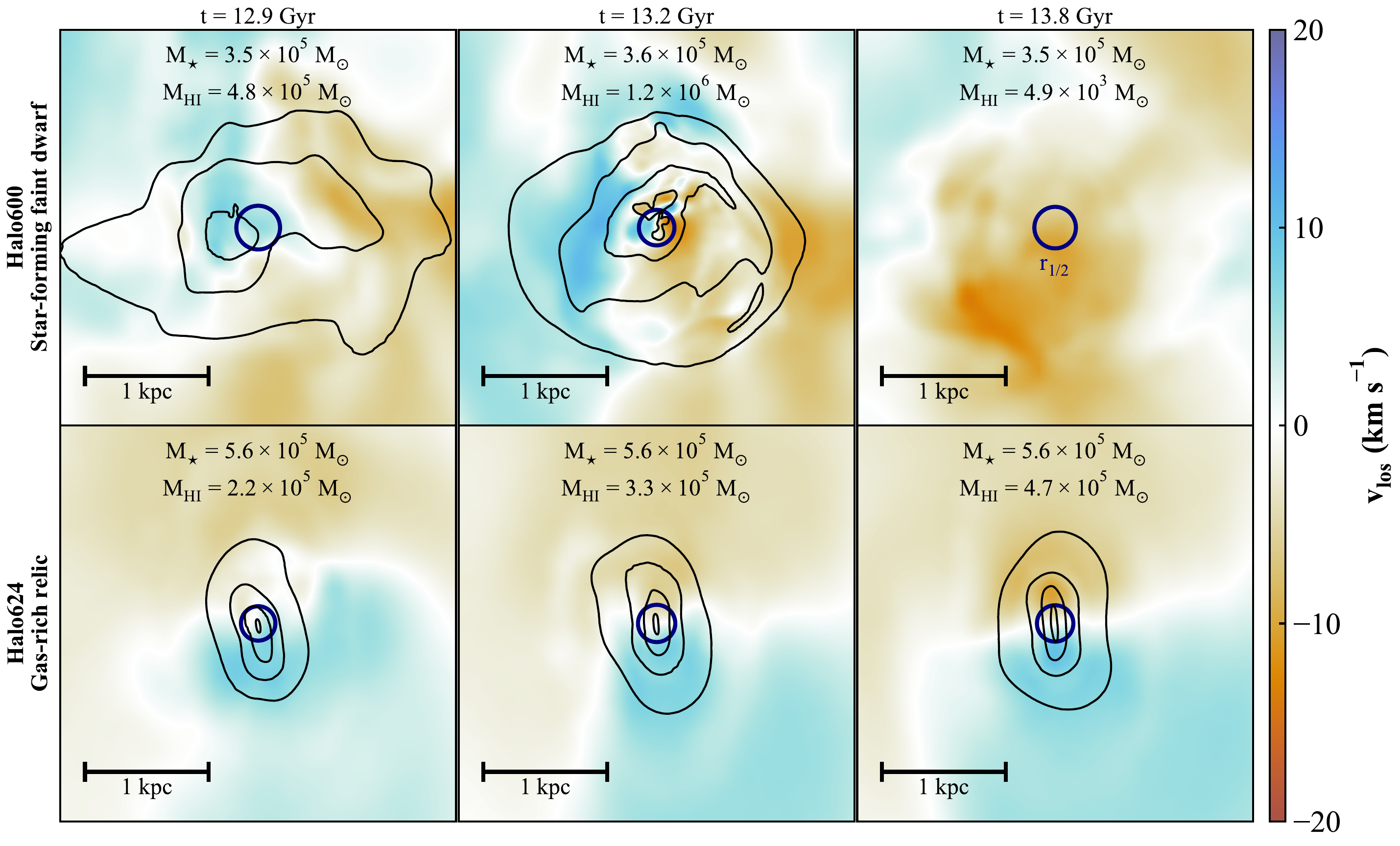}

    \caption{Maps of the gas velocity, \hinospace-mass weighted along the same line of sight for a star-forming (top) and quiescent (bottom) simulated dwarf. From left to right, we show the galaxies at three different times. Actively star-forming dwarfs (top) show irregular and time-varying \hi distributions (the black contours show $10^{18}$, $10^{19}$, $10^{20}$ and $10^{21} \, \cmsquare$ column densities) following their cycle of gas accretion and stellar feedback (\citealt{Rey2022EDGEHI}). This dynamic behaviour is reflected in their gas kinematics, which are often disturbed and show rotational support in short-lived episodes (e.g. top, centre; Figures~\ref{fig:vphitimeevolution} and~\ref{fig:shortliveddisc}). In contrast, galaxies with quieter histories (e.g. bottom is yet to reignite star formation after quenching at $z=4$) show stable \hi reservoirs within their half-light radius (blue circles) that can host long-lived, stable \hi discs (Figure~\ref{fig:rotationcurve}).}
    \label{fig:hivelmaps}
\end{figure*}

Figure~\ref{fig:hivelmaps} illustrates the diversity of gas and \hi kinematics found across our sample of simulated gas-rich faint dwarfs. The top panels show an actively star-forming dwarf; the bottom a quiescent dwarf; each at three output times spanning the last billion years of their evolution (selected to be the same output times as in \citealt{Rey2022EDGEHI}, fig.~1). The maps plot the gas velocity weighted by \hi mass along the same line of sight in all panels. Contours show constant $10^{18}$, $10^{19}$, $10^{20}$ and $10^{21} \, \cmsquare$ \hi column densities (black) and $\rhalflight$ (blue circles).

Focusing first on the star-forming dwarf (top panels), we recover \hi distributions that are strongly and rapidly varying in time (see \citealt{Rey2022EDGEHI} for an in-depth quantification). In these low-mass systems, stellar feedback drives asymmetric and disturbed \hi morphologies (top, middle), that are often offset from the galaxy's stellar distribution (e.g. top, left), and that can become temporarily unobservable following powerful outflow and heating (top right; $\Mhi$ marked on each panel). This dynamic and time-varying behaviour is reflected in the gas kinematics. Over one billion years, gas travels in bulk flows of opposite directions within $\rhalflight$ (top, left and right), but also exhibits a snapshot of potential rotation with a disturbed but apparent gradient in line-of-sight velocity around $\rhalflight$ (top, centre).

In contrast, the quiescent system (bottom row) shows much more stable gas content, slowly accumulating \hi gas over the last billion years (growing $\Mhi$ at constant $\Mstar$; see also \citealt{Rey2020, Rey2022EDGEHI}).  Furthermore, this \hi reservoir shows a distinctly flattened morphology, with a positive-to-negative line-of-sight velocity gradient across $\rhalflight$ at all time stamps.

These two examples summarize well the more complete and quantitative investigation presented in the next section. Star-forming low-mass dwarfs host short-lived instances of \hi rotation across their evolution, but the small and tenuous discs are rapidly disrupted by the energy input from newborn massive stars. Quiescent systems have more stable \hi reservoirs and kinematics, increasing (but not guaranteeing) their chances to host organized and long-lived \hi rotation ideal for inferences of the structure of their host dark matter halo.

\section{Rotating \hi discs in faint dwarfs and their physical drivers} \label{sec:physicaldrivers}

\begin{figure*}
  \centering
    \includegraphics[width=\textwidth]{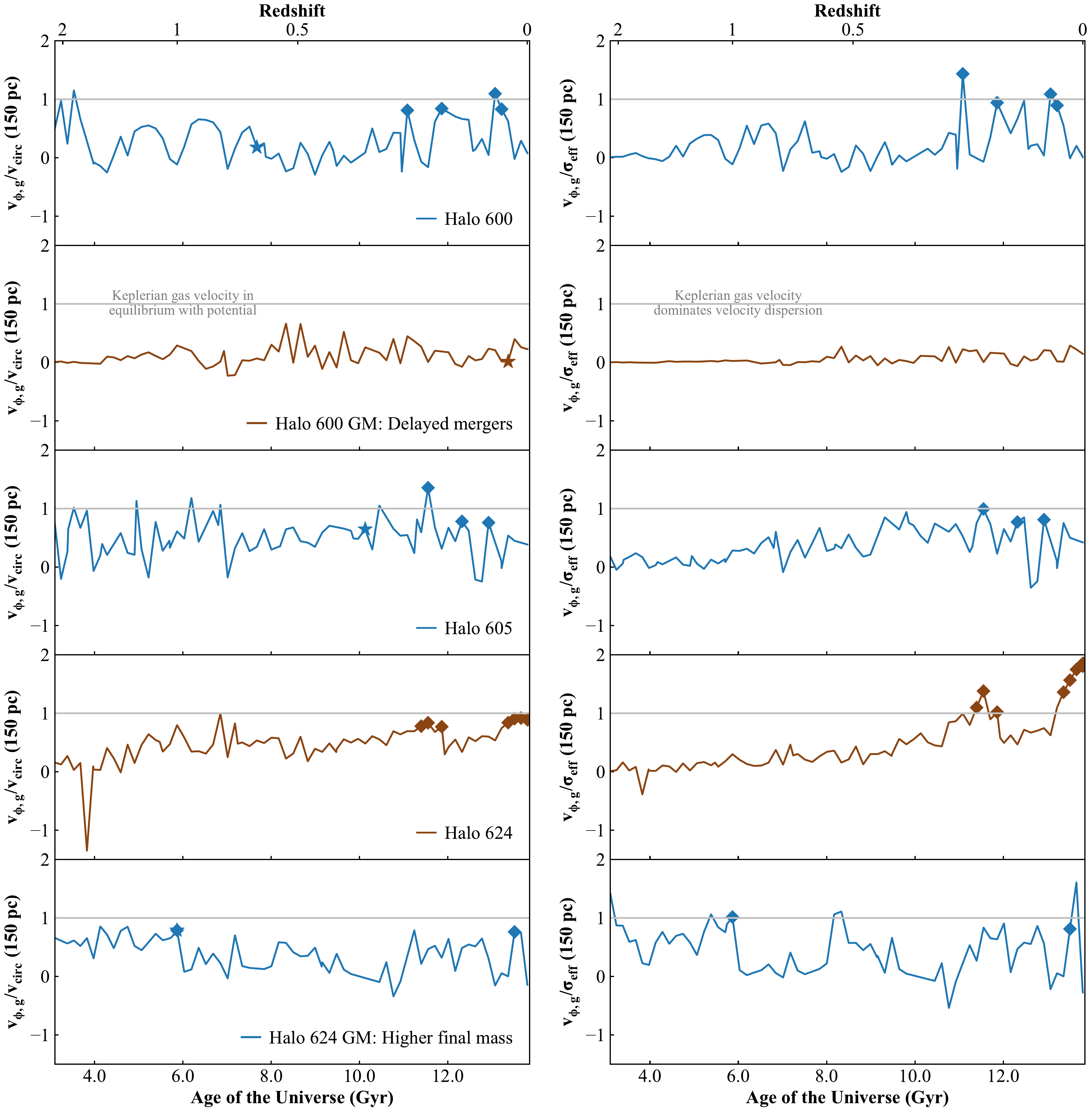}

    \caption{Time evolution of $\vphi / \vcirc$ (left; i.e. a proxy for gas rotation in equilibrium with the gravitational potential) and $\vphi / \vdispeff $ (right; i.e. comparing rotational-to-pressure support) across our suite of simulated \hinospace-bearing dwarfs after reionization. Star-forming galaxies (blue, stars marking their first time of post-reionization star formation) show instances where they host rotationally-supported \hi kinematics close to equilibrium with the gravitational potential ($\vphi \approx \vcirc$ and $\vphi \geq \vdispeff$; marked with diamonds). These episodes are short-lived, as the \hi is rapidly heated and dispersed by stellar feedback (Figure~\ref{fig:shortliveddisc}). Quiescent galaxies (brown lines) have more stable kinematics, with one hosting long-lived, stable \hi rotation that provides an ideal target for dark matter inferences (Figure~\ref{fig:rotationcurve}). 
    }
    \label{fig:vphitimeevolution}

\end{figure*}

We now aim to gain more quantitative insights into the gas rotational support of our galaxies. In particular, we wish to (i) establish whether organized gas rotation can dominate thermal and turbulent motions, and thus be clearly identified observationally; (ii) test whether this rotation is close to circular and in equilibrium, and thus easy to relate to the host gravitational potential; and (iii) gain insights into the prospects of characterizing such rotating gas with radio interferometers.

Gas contents and kinematics in our galaxies can be strongly varying on time-scales comparable to the local dynamical times and to the lifetime of SNII (both $\approx 10 \, \Myr$), particularly for actively star-forming objects. Unfortunately, this time-scale is much shorter than the cadence with which we save simulation outputs for each galaxy ($\approx 100\, \Myr$). Combined with the Eulerian nature of our simulation code that limits the tracking of gas over time, this makes it challenging to establish causal evolutionary trends from one snapshot to the next (e.g. in Figure~\ref{fig:hivelmaps}, the top panels are difficult to relate to one another as multiple star formation bursts and \hi re-accretion has occurred between them). Similarly to \citealt{Rey2022EDGEHI}, we thus adopt a statistical approach, treating each snapshot as an independent realization of the star forming cycle, flagging times of potential organized gas rotation across each dwarf's history in subsection~\ref{sec:sec:vphiovertime}, and then study the \hi kinematics at those times in more details in subsections~\ref{sec:sec:shortlived} and~\ref{sec:sec:longlived}. In future work, we will alleviate this issue leveraging algorithms that allow high-cadence tracking of gas dynamics (\citealt{Cadiou2019}) to connect star formation activity and \hi properties more causally (S. Hutton et al. in preparation).

\subsection{Existence and prevalence of gas rotation} \label{sec:sec:vphiovertime}

We start by computing, for each simulated snapshot, profiles of the tangential gas velocity $\vphi$, the circular velocity $\vcirc$, the 3D isothermal sound speed $c_s$ and the 3D gas turbulent velocity $\vdisp$ to quantify rotational, gravitational, thermal, and turbulent support respectively (see Appendix~\ref{app:pressuresupport} for formal definitions). We then compute the projected radial profiles of $\vphi$, $c_s$ and $\vdisp$ viewed face-on (i.e. in the plane of the disc) in 100 bins linearly spaced between 0 and 2 kpc and construct the effective velocity dispersion of the gas $\vdispeff= \sqrt{c_s^2 + \vdisp^2}$. The 3D $\vcirc$ profile is derived from the full gravitational potential in the same radial range, sourced by the combination of dark matter, gas and stars (but strongly dominated by the dark matter at all radii for these faint objects).

Figure~\ref{fig:vphitimeevolution} shows the evolution of $\vphi / \vcirc$ and $\vphi / \vdispeff$ evaluated at $150 \, \pc$, where the highest column density \hi is most often found (\citealt{Rey2022EDGEHI}). We only show the late-time evolution of these dwarfs ($z\leq2$), that is when they host detectable \hi (see \citealt{Rey2022EDGEHI}, fig.~1 for the time evolution of $\Mhi$ over time of each of these galaxies) and omit their earlier phase where saved simulation outputs are sparser and miscentring due to mergers make $\vphi$ and $\vcirc$ even noisier. Trends in Figure~\ref{fig:vphitimeevolution} are qualitatively unchanged if measuring velocities at $100$, $200 \, \pc$ or at each galaxy's $\rhalflight$ instead.

Focusing first on $\vphi / \vcirc$ (left-hand panels), we recover that gas kinematics are strongly variable in time, without clear evolutionary trends for star-forming low-mass dwarfs (blue) after the re-ignition of their star formation (marked by stars in Figure~\ref{fig:vphitimeevolution}). This is expected as stellar feedback efficiently disrupts the ISM in these shallow potential wells ($\vcirc \approx 10 \, \kmpers$ at $\rhalflight$). Nonetheless, some peaks approach $\vphi / \vcirc \approx 1$ (grey line in Figure~\ref{fig:vphitimeevolution}) indicating potential short-lived episodes where the rotational velocity $\vphi$ is close to equilibrium with $\vcirc$ sourced by the underlying gravitational potential. Furthermore, during these episodes, rotational motions can dominate over turbulent support, with $\vphi / \vdispeff \geq 1$ (right-hand panels). 

To quantify this further, we extract all time instances when $0.75 \leq \vphi / \vcirc  \leq 1.25$ (i.e. loosely bracketing gas in circular rotation, acknowledging that $\vphi$ is not yet corrected for pressure support; see Section~\ref{sec:sec:shortlived} and Appendix~\ref{app:pressuresupport}) and $\vphi / \vdispeff \geq 0.75$ (i.e. rotation loosely dominating over thermal and kinetic turbulence). These cuts should not be interpreted quantitatively, but rather as helpful to flag the likely presence of a galactic gas disc in the noisy kinematics of our sensitive objects. We mark these times with diamonds in Figure~\ref{fig:vphitimeevolution}, finding at least two examples satisfying these conditions per star-forming dwarf. As we will see in Section~\ref{sec:sec:shortlived}, these snapshots can showcase interpretable but short-lived \hi rotation curves. 

\begin{figure*}
  \centering
    \includegraphics[width=\textwidth]{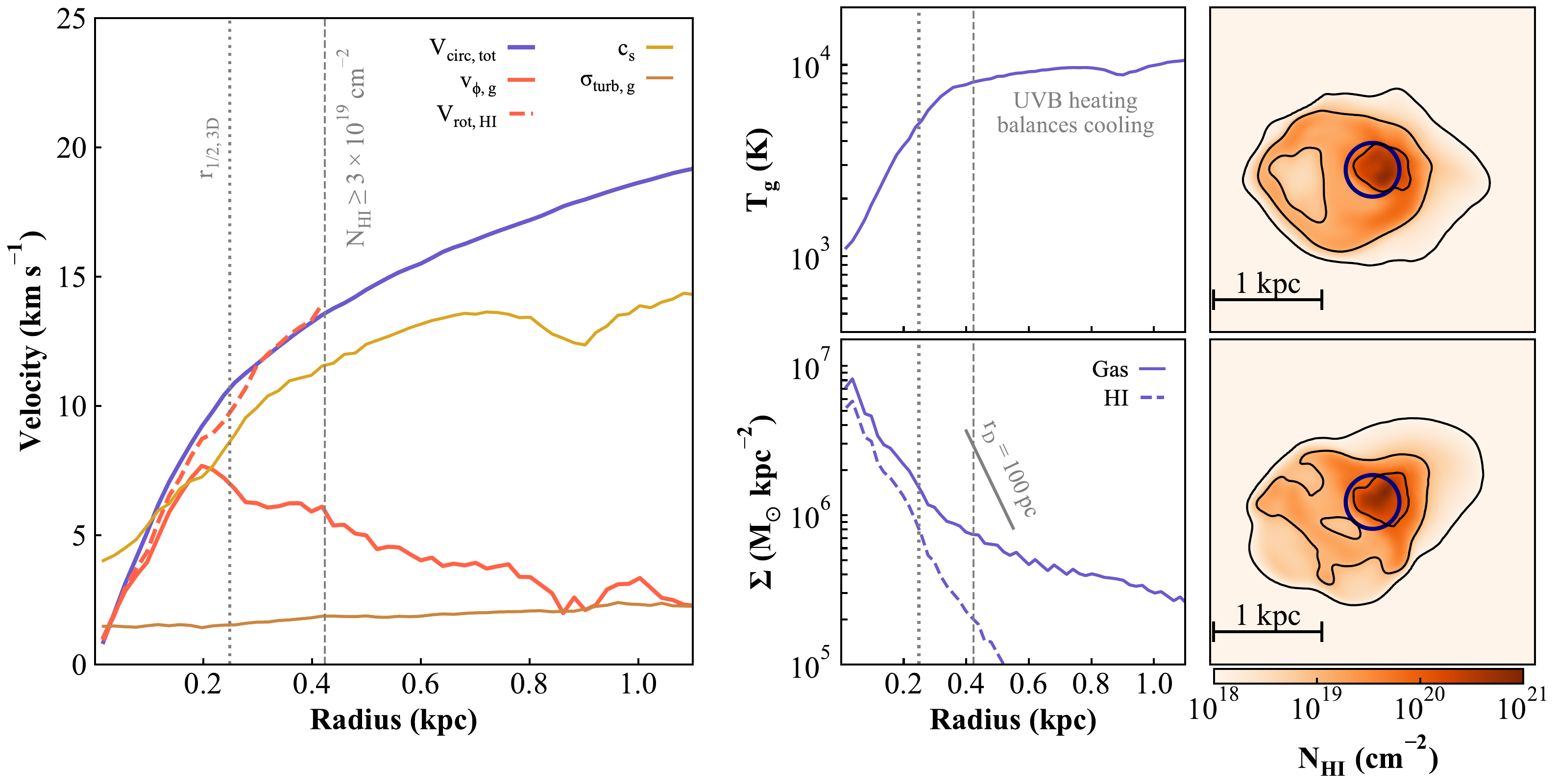}

    \caption{\hi kinematics in an example star-forming \hinospace-rich low-mass dwarf galaxy (`Halo 600') singling a time of ordered \hi rotation ($t=11.9\, \Gyr$). The face-on 2D tangential velocity profile of the gas (left, red) follows the rise of the 3D rotation curve (left, blue) sourced by the gravitational potential, with the \hi distribution extending to $\approx 2 \rhalflight$ and showcasing a cold and close-to-exponential disc structure (middle panels; indicative exponential scale lengths in grey). Stellar feedback drives asymmetric \hi features in the outskirts (right panels, face-on and side-on projections in the top and bottom panels, respectively), and photo-heating from the UV background leads to a rising thermal support (left, $c_s$ in gold and top, middle). Despite these features, traditional pressure-support corrections to the gas velocity (dashed red; also called asymmetric drift, see Appendix~\ref{app:pressuresupport}) can accurately recover $\vcirc$ out to $\NhiaboveXX{3}{19}$.
    }
    \label{fig:shortliveddisc}

\end{figure*}

Contrasting again with our star-forming examples, quiescent dwarfs (brown lines in Figure~\ref{fig:vphitimeevolution}) show more stable evolution over time and clearer evolutionary trends. One galaxy (second row) lacks evidence for gas rotation ($\vphi / \vcirc \approx 0$) at all times, but the other (fourth row; also bottom panels of Figure~\ref{fig:hivelmaps}) is regularly approaching $\vphi / \vcirc \approx 1$, and $\vphi / \vdispeff \geq 1$, and notably over its entire last billion years of evolution. As we will see in Section~\ref{sec:sec:longlived}, this galaxy hosts a stable, long-lived \hi disc with an easy-to-interpret rotation curve. 

Our analysis thus shows that intrinsic ordered gas rotation should be expected in low-mass, \hinospace-bearing dwarfs. However, stellar feedback in star-forming objects can efficiently disrupt small gas discs ($\vcirc \approx 10 \, \kmpers$), making them short-lived and rare. This provides a natural explanation for the lack of observed rotation in the faintest dwarfs (e.g. \citealt{Bernstein-Cooper2014, Adams2018, McQuinn2021}). Quiescent, \hinospace-bearing dwarfs, that are yet to re-ignite star formation after cosmic reionization, offer a contrastingly calmer and more stable environment. This promotes well-ordered and long-lived gas rotation, with greater prospects for dark matter science using rotation curves which we quantify next. 

\subsection{Short-lived \hi discs in star-forming low-mass dwarfs} \label{sec:sec:shortlived}

To quantify \hi kinematics in the noisy, star-forming dwarfs, we visually inspect individual rotation curves and \hi column density maps at the times flagged to have higher probabilities of galactic gas discs (Figure~\ref{fig:vphitimeevolution}, diamonds). Appendix~\ref{app:shortliveddiscs} presents the full results of this systematic inspection, showcasing very diverse \hi distributions with complex spatial, kinematic and thermodynamical structures due to stellar feedback. 

But Figure~\ref{fig:shortliveddisc} (and other examples in Appendix~\ref{app:shortliveddiscs}) show that, even if short-lived, organized \hi rotation can occur in these systems. Figure~\ref{fig:shortliveddisc} shows the total gas velocity profiles (left panel), surface density and temperature profiles (middle panels) and \hi column density maps viewed face-on and edge-on (right panels) of `Halo 600' at $t=11.9\, \Gyr$. At this time, the \hi distribution is spatially extended (right panels), reaching $\NhiaboveXX{3}{19}$ outside  $\rhalflight$ (dotted and dashed lines in left panel). Such surface brightnesses are at the limit of what can be achieved by deep follow-ups with current-generation interferometers in faint dwarfs (e.g. \citealt{Adams2018}). Furthermore, despite showcasing holes and being lopsided at large radii, the \hi distribution is smooth and regular in the inner galaxy. 

In fact, the $\vphi$ profile (left, red) follows the rise of $\vcirc$ (blue) within $200 \, \pc$, as expected from equilibrium circular orbits. The \hi gas also exhibits a close-to-exponential radial surface brightness profile (middle, bottom), reminiscent of classical rotation curves of galactic discs. However, rotation only marginally dominates compared to the primary source of gas velocity dispersion (thermal pressure, $c_s$ in gold) and only at specific radii. Extracting and claiming a rotational signal from moment maps will thus be challenging once observational challenges associated with such faint and small objects are folded-in (discussed further in Section~\ref{sec:conclusion}). Nonetheless, computing the standard pressure correction to $\vphi$ (also called asymmetric drift; see Appendix~\ref{app:pressuresupport} for further details), we obtain the \hi rotational velocity ($\vrothi$; red, dashed) which accurately recovers $\vcirc$ deep into the diffuse \hi regime ($\NhiaboveXX{3}{19}$; $\approx 2 \rhalflight$). These results are highly promising and show that, although rare and potentially difficult to identify, \hi rotation curves can be harnessed for dark matter science in star-forming low-mass dwarfs. 

Extending this analysis further is complicated by the unusual thermal structure and density profile of the gas compared to higher-mass galaxies. The temperature is steadily rising when moving to the outskirts (top, middle), with $c_s$ following accordingly. Already at $\approx 2\, \rhalflight$, thermal pressure fully dominates rotational signals and \hi has transitioned from colder ($T\approx 10^3\, \K$) to warmer temperatures ($\approx 10^4\, \K$). This transition also materializes in a change of slope of the gas surface density profile (middle, bottom). 

This structure is naturally explained by the rising importance of the cosmic ultraviolet background (UVB) at low-galactic masses. Following cosmic reionization, the UVB provides a source of ionization and heating that maintains diffuse gas in photo-ionization equilibrium around $T \approx 10^4\, \K$. Galaxies considered here have potential wells just deep enough to accrete fresh gas from their diffuse surroundings (e.g. $c_s$ is only slightly below $\vcirc$ at large radii). Gas can self-shield and cool below $10^4 \, \K$ in the centre of the dwarf (\citealt{Rey2020}) but gas in the outskirts rapidly transitions to $\approx10^4 \, \K$ in a balance between gravity, cooling from metal lines and photo-heating from the UVB (e.g. \citealt{Ricotti2009, Rey2020,Benitez-Llambay2020}). The detection of warm \hi ($10^4 \, \K$) in projection thus cannot be unequivocally attributed to photo-heating from stars in these faint objects, particularly when undertaking deep observations probing the diffuse gas (e.g. \citealt{Adams2018}). 

Establishing and interpreting \hi rotation is likely to be challenging in star-forming faint dwarfs. Although one can recover the gravitational potential with access to all simulated information to compute thermal support, how to achieve this feat from \hi datacubes is less clear. The dominance of pressure terms over rotation might point to the need to introduce new approaches to infer dark matter profiles (e.g. starting from hydrostatic equilibrium rather than axisymmetric rotation; \citealt{Patra2018}). Furthermore, this also yields thicker \hi discs compared to more massive disc galaxies. Once viewed inclined, the \hi linewidth from thicker discs receives contribution along the line of sight, leading to a potential mismatch between the rotation velocity measured from the \hi and the intrinsic value. And even if clear rotation can be established, the sensitivity of these galaxies to stellar feedback makes a detailed assessment of rotation curve systematics essential for robust dark matter inferences (see Appendix~\ref{app:shortliveddiscs} for examples of out-of-equilibrium flows, non-circular motions, feedback-driven holes and \citealt{Read2016DwarfRCs, Oman2019, Downing2023} for further discussion). 

Performing these quantifications would be best undertaken by generating mock \hi datacubes from our simulated snapshots to assess the robustness of standards rotation curve fitting methods (e.g. \textsc{3dbarolo}; \citealt{DiTeodoro2015}) and understand whether new approaches would be better suited to recover dark matter information. We are currently developing a package that can easily incorporate different pressure terms, and treat the impact of disc thickness on the line-of-sight velocity distribution, and leave the quantifications of these uncertainties to future work. In the next Section, we instead focus on easier-to-interpret and long-lived \hi rotation curves that can be found in quiescent dwarfs. 

\subsection{Long-lived \hi discs in quiescent dwarfs} \label{sec:sec:longlived}

\subsubsection{Circular, equilibrium \hi rotation in a low-mass dwarf}

\begin{figure*}
  \centering
    \includegraphics[width=\textwidth]{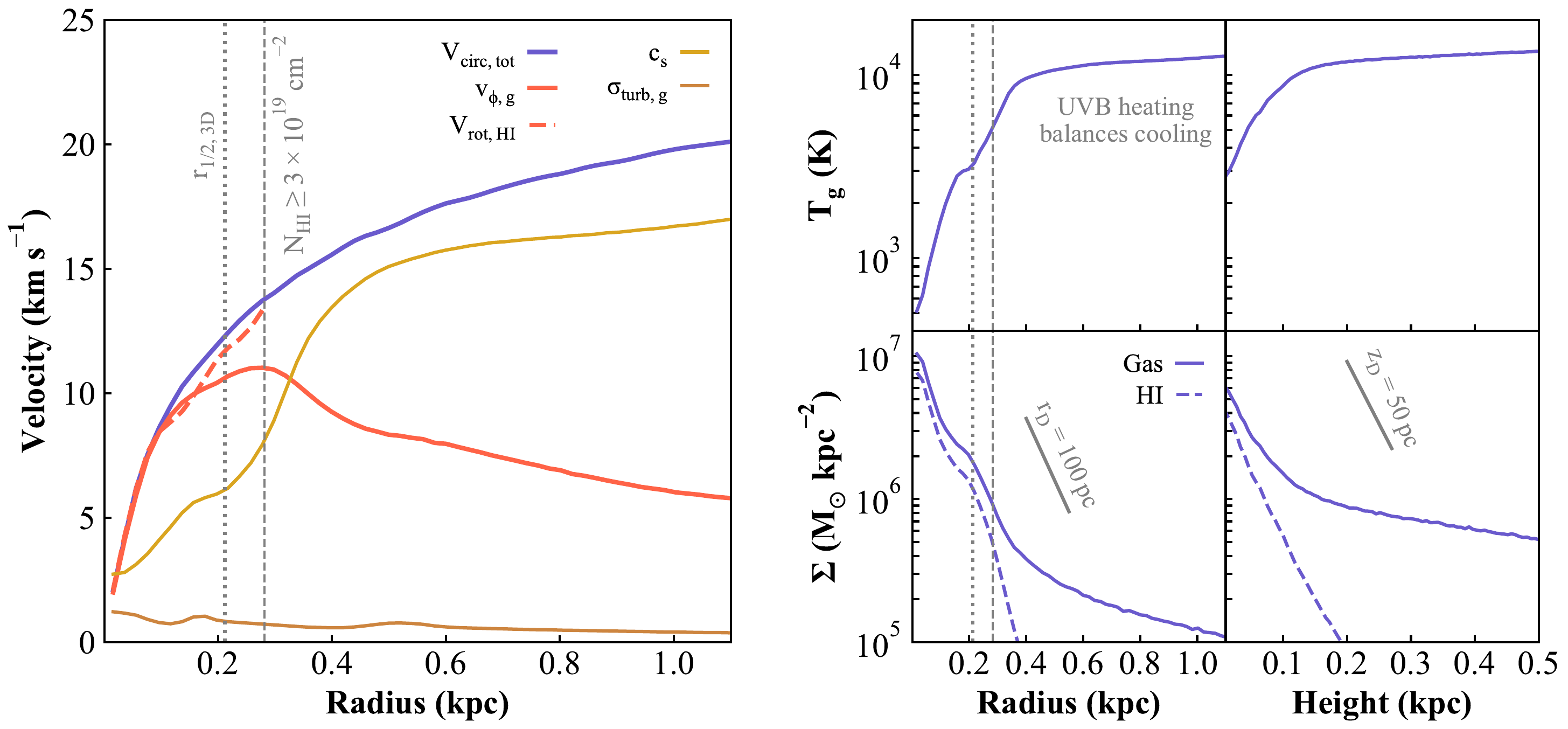}

    \caption{Classical \hi disc at $z=0$ in a quiescent low-mass dwarf (`Halo 624'). The tangential velocity profile (left, red) accurately tracks the rise of the rotation curve (blue) within $\rhalflight$ (dotted). Correcting the tangential velocity for pressure support (dashed) helps modelling the rising thermal support (gold) and recover $\vcirc$ further into the outskirts of the \hi distribution (dashed shows $\NhiaboveXX{3}{19}$) where the disc gets thicker (right panels). This long-lived and easy-to-interpret \hi rotation curve is unique to this object, driven by the defining shape of the host dark matter halo (Figure~\ref{fig:haloshape} and~\ref{fig:angmomorientation}). Characterizing such a rotation curve would prove invaluable to obtain robust inferences of inner dark matter density profiles.
    }
    \label{fig:rotationcurve}

\end{figure*}

Figure~\ref{fig:rotationcurve} shows the rotation curve at $z=0$ (left panel), the gas temperature, and surface density radial and vertical profiles (right panels) for the quiescent dwarf hosting a clear, long-lived rotation signal (Figure~\ref{fig:vphitimeevolution}, fourth row). 

In this example, $\vphi$ accurately tracks $\vcirc$ without corrections to $\approx 10 \, \kmpers$, indicating near-perfect circular rotation in equilibrium with the gravitational potential. Compared to our star-forming example (Figure~\ref{fig:shortliveddisc}), the inner gas is cold (right, top panels), and rotation strongly dominates thermal support and turbulence ($c_s$ and $\vdisp$ in gold and brown) in the inner galaxy. Pressure corrections (red, dashed) are subdominant at all radii, only becoming significant when reaching more diffuse \hi ($ \geq \rhalflight$; dotted line) brought to warmer temperatures by the UVB. 

The lack of disturbances from star formation in this object also ensures a regular, symmetric and well-ordered \hi distribution (recall Figure~\ref{fig:hivelmaps}), showcasing close-to-exponential \hi radial and vertical profiles (bottom panels). The vertical profile (right panels) shows a thickened \hi disc (aspect ratios approaching 1:2; indicative exponential scale lengths in grey), as expected from the rising importance of pressure support towards larger radii. 

To summarize, this quiescent galaxy hosts a classical \hi rotation curve, at column densities achievable by current-generation radio interferometers ($\NhiaboveXX{3}{19}$, dashed-grey vertical line). This rotation curve is comparatively easy to interpret, holding great promise for extracting unbiased estimates of the inner dark matter density profiles. Even further, we show in Appendix~\ref{app:longliveddisc} that a similar rotation structure is present over the last two billion years of evolution of this galaxy (see also diamonds in Figure~\ref{fig:vphitimeevolution}), with the cold and circular \hi rotation curve being in place for the last $500 \, \Myr$. The excellent agreement between the \hi rotation and the gravitational potential is thus long-lived and little disrupted.

Our results strongly motivate targeting low-mass dwarfs with quieter evolutions when searching for high-quality \hi rotation curves. Such quiescent candidates have already been reported (e.g. \citealt{Janesh2019}) and their follow-up with deep and high-resolution \hi interferometers should be given high priority. However, our analysis also shows that a lack of star formation activity is insufficient to guarantee well-behaved rotation curves -- the other quiescent galaxy in our suite does not exhibit rotation (second row in Figure~\ref{fig:vphitimeevolution}) and star-forming examples lack clear signals during their quiescent periods (notably before the re-ignition of their star-formation marked by a star in Figure~\ref{fig:vphitimeevolution} ending several billion years of quiescent evolution). We thus turn next to understanding what leads to long-lived \hi discs in this specific object.

\subsubsection{The link between \hi rotation and the shape of the host dark matter halo} \label{sec:sec:shape}

\begin{figure}
  \centering
    \includegraphics[width=\columnwidth]{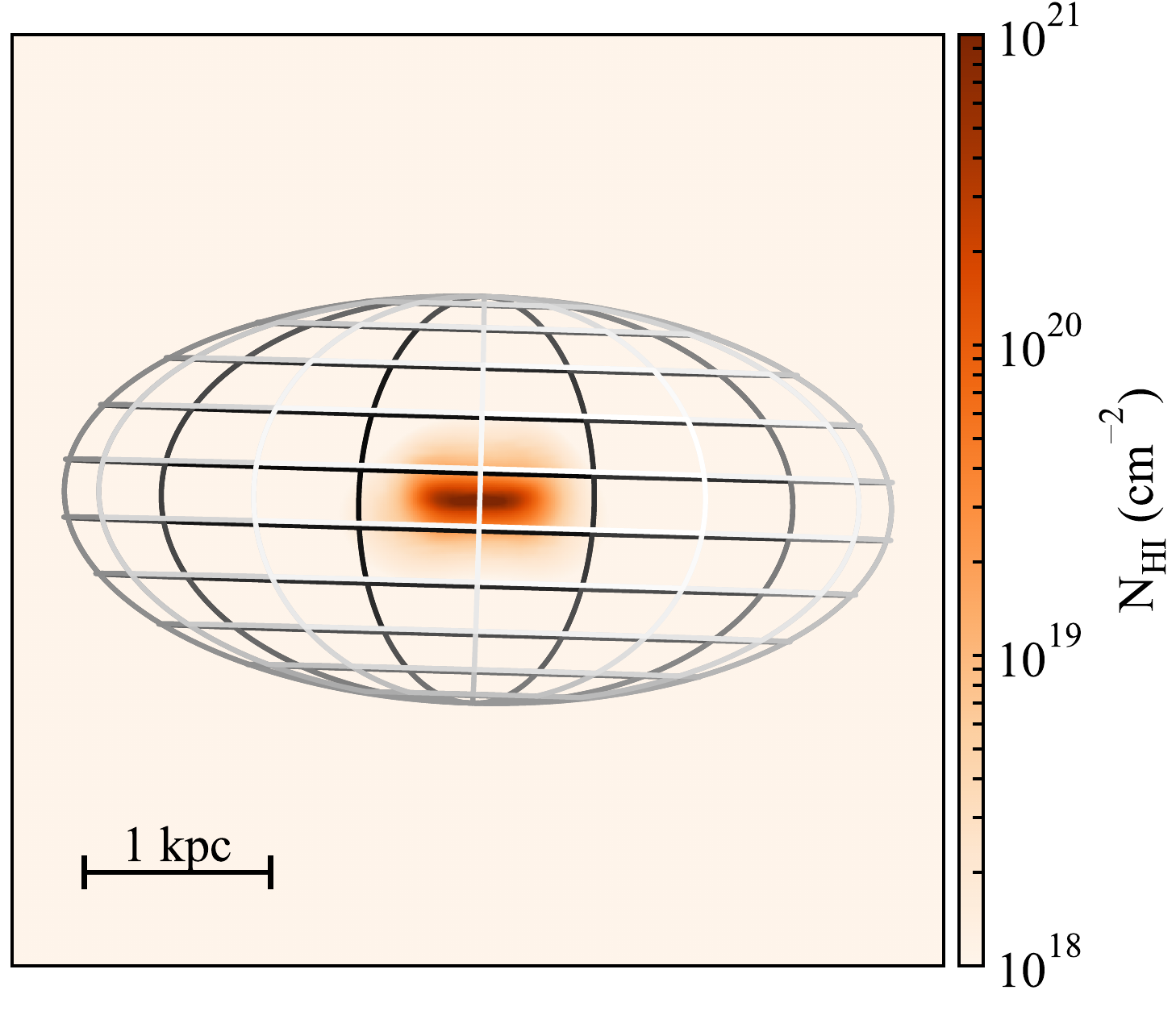}

    \caption{\hi column density map oriented edge-on with respect to the gas for the galaxy hosting stable rotation (same as Figure~\ref{fig:rotationcurve}). The host dark matter halo is strongly oblate (white-grey-black mesh showcasing the 3D shape), defining an axisymmetric geometry well aligned with the revolution axis of the \hi disc. This configuration induces torques that align infalling gas into the plane of the \hi disc (Figure~\ref{fig:angmomorientation}) and favour its growth.}
    \label{fig:haloshape}

\end{figure}

A defining feature of our quiescent galaxy with long-lived \hi disc lies in the properties of its host dark matter halo. The specific merger history of this object, particularly a major interaction at $z\approx4$ leads to a strongly oblate dark matter halo shape compared to the more triaxial or prolate shapes across the rest of the simulated suite ($b/a \approx0.9$, $c/a \approx 0.5$ between $r\approx200\, \pc$ and $r\approx 20 \, \kpc$ for this halo; see \citealt{Orkney2023}, fig. 1). We link these two aspects in Figure~\ref{fig:haloshape}, visualizing the alignment between the \hi angular momentum compared to the halo shape. We plot the \hi column density map at $z=0$, oriented side-on compared to the angular momentum of gas with $\xhi \geq 0.5$ and overlay the 3D halo shape computed exclusively from the dark matter particles as in \citet{Orkney2023} (grey mesh; whiter towards the foreground, blacker towards the background). Note that \citet{Orkney2023} derive halo shapes using higher-resolution re-simulations ($\mdm = 120 \, \Msol$) of the galaxies studied in this work -- we have checked that (i) the radial profile of axis ratios at $r\geq 100\, \pc$ and (ii) the presence and orientation of the gas disc at $z=0$ are both consistent between the two resolutions. This also validates that the presence and formation of the \hi disc is physical, rather than stochastic or resolution-limited. 

The \hi disc and the flattened axis of the oblate dark matter halo are exceptionally well aligned in Figure~\ref{fig:haloshape}. This is best understood by the naturally axisymmetric geometry of a significantly oblate halo. Such geometry induces torques that align accreting gas along its revolution axis, a process best studied in the case of axisymmetric torques induced by galactic stellar discs (see e.g. \citealt{Danovich2015} for a discussion). Here, these torques are sourced by the dark matter halo itself, as the gas and stars contribute only marginally to the gravitational potential.

\begin{figure}
  \centering
    \includegraphics[width=\columnwidth]{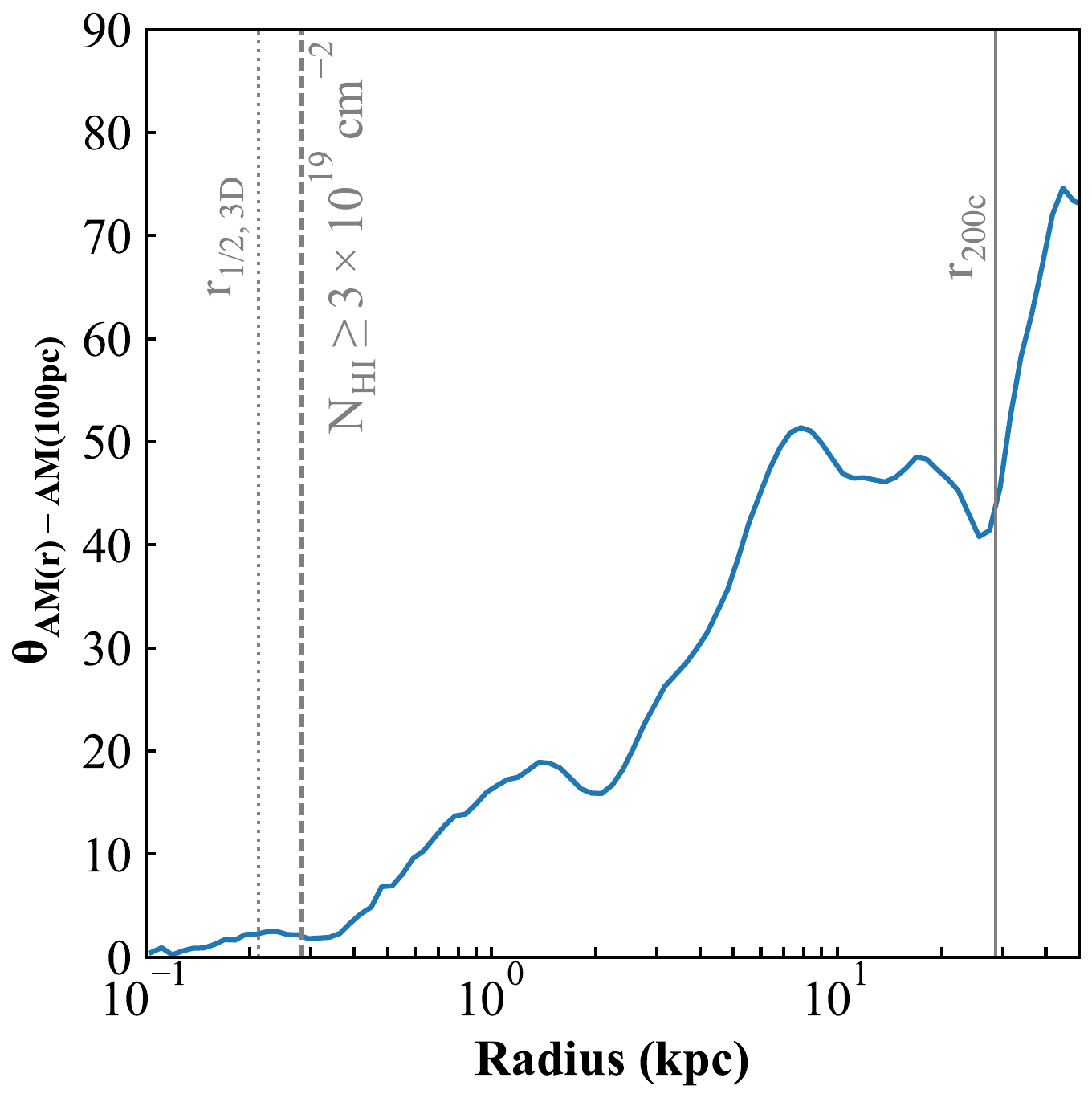}

    \caption{Orientation of the angular momentum of the gas in a radial shell compared to that in the inner 100 pc, for the same galaxy as in Figure~\ref{fig:rotationcurve}. Gas outside the virial radius (grey line) is accreted tilted compared to the inner angular momentum ($\theta \leq 50^{\circ}$), but is gradually and coherently torqued with decreasing radius to align with the inner \hi disc ($\theta\approx0$ for $r\leq 300\, \pc$). The revolution axis of the \hi disc also coincides with that of the dark matter halo shape (Figure~\ref{fig:haloshape}).
    }
    \label{fig:angmomorientation}

\end{figure}

To visualize this torque in action, Figure~\ref{fig:angmomorientation} shows the orientation of the gas angular momentum in a given radial shell compared to the angular momentum of the gas in the inner 100 pc (which is almost purely \hinospace; Figure~\ref{fig:rotationcurve}). Starting from outside the virial radius ($\geq 30 \, \kpc$), gas is accreted with significant angular momentum but orthogonal to the inner disc ($\theta \geq 50^{\circ}$) before stabilizing around this angle across between 8 and 20 kpc. Towards smaller radii, however, the gas gradually gets torqued to align with the inner angular momentum of the galaxy ($\theta \leq 10^{\circ}$ within $\approx 3 \rhalflight$), at which point it shares the same revolution axis as that of the oblate dark matter halo shape (Figure~\ref{fig:haloshape}). This gradual realignment of gas throughout the halo, starting at radii well outside the galaxy, firmly establishes the causal link between the halo shape and the presence of the \hi disc. 

\section{Summary and discussion} \label{sec:conclusion}

We have analysed the gas and \hi kinematics of simulated low-mass ($10^4 \leq \Mstar \leq \xMsol{2}{6}$) dwarf galaxies, first introduced in \citet{Rey2019UFDScatter, Rey2020} and evolved to $z=0$ using high-resolution ($\approx 3\, \pc$) zoomed cosmological simulations using the \textsc{edge} galaxy formation model (\citealt{Agertz2020EDGE}). We studied five dwarf galaxies that are close analogues to the observed population of faint, but gas-rich and \hinospace-bearing dwarfs ($10^5 \leq \Mhi \leq 10^6 \, \Msol$; see \citealt{Rey2022EDGEHI} for a more detailed comparison the observed population; \citealt{Irwin2007, Cole2014}; \citealt{McQuinn2015, McQuinn2020, McQuinn2021}; \citealt{Sand2015, Adams2018, Brunker2019, Janesh2019, Hargis2020, Bennet2022, Rhode2023}). 

At this galactic mass-scale, galaxy formation effects within $\Lambda$CDM are inefficient at dynamically heating dark matter into flat and large ($\approx \rhalflight$) dark matter cores, leading to steep dark matter density profiles at $\rhalflight$ in all of our dwarfs (Figure~\ref{fig:dmdensity_profiles}). Inferring the structure of dark matter haloes in this regime, for example through \hi rotation curves, thus holds great promise to pinpoint the relative contributions of dark matter microphysics and galaxy formation in driving dark matter heating. 

We find that simulated low-mass dwarfs that are actively forming stars undergo strong variability in their \hi distributions, driven by the cycle of gas accretion and efficient stellar feedback (\citealt{Rey2022EDGEHI}). This variability is reflected in their \hi kinematics, showcasing disturbed and rapidly changing gas flows (Figure~\ref{fig:hivelmaps}) as supernovae easily disrupt gas dynamics in these shallow potential wells ($\vcirc$ and $\vphi \approx 10 \, \kmpers$ at $\rhalflight$). We find occasional, short-lived ($\ll 150 \, \Myr$) episodes of organized \hi rotation in these star-forming objects (Figure~\ref{fig:vphitimeevolution}), for which rotation curves can recover the underlying gravitational potential (Figure~\ref{fig:shortliveddisc}). But the prevalence of out-of-equilibrium feedback-driven gas flows and the (comparatively) high velocity dispersions due to thermal support ($\vdispeff \approx 10 \, \kmpers$) lead to difficult-to-interpret rotation curves (see also Appendix~\ref{app:shortliveddiscs}). Clear and robust \hi rotation that can be harnessed for dark matter science is thus expected to be rare in these active systems, aligning with the lack of observed rotation in the handful of low-mass star-forming dwarfs with detailed \hi observations (e.g. \citealt{Bernstein-Cooper2014, Adams2018, McQuinn2021}).

\begin{figure*}
  \centering
    \includegraphics[width=\textwidth]{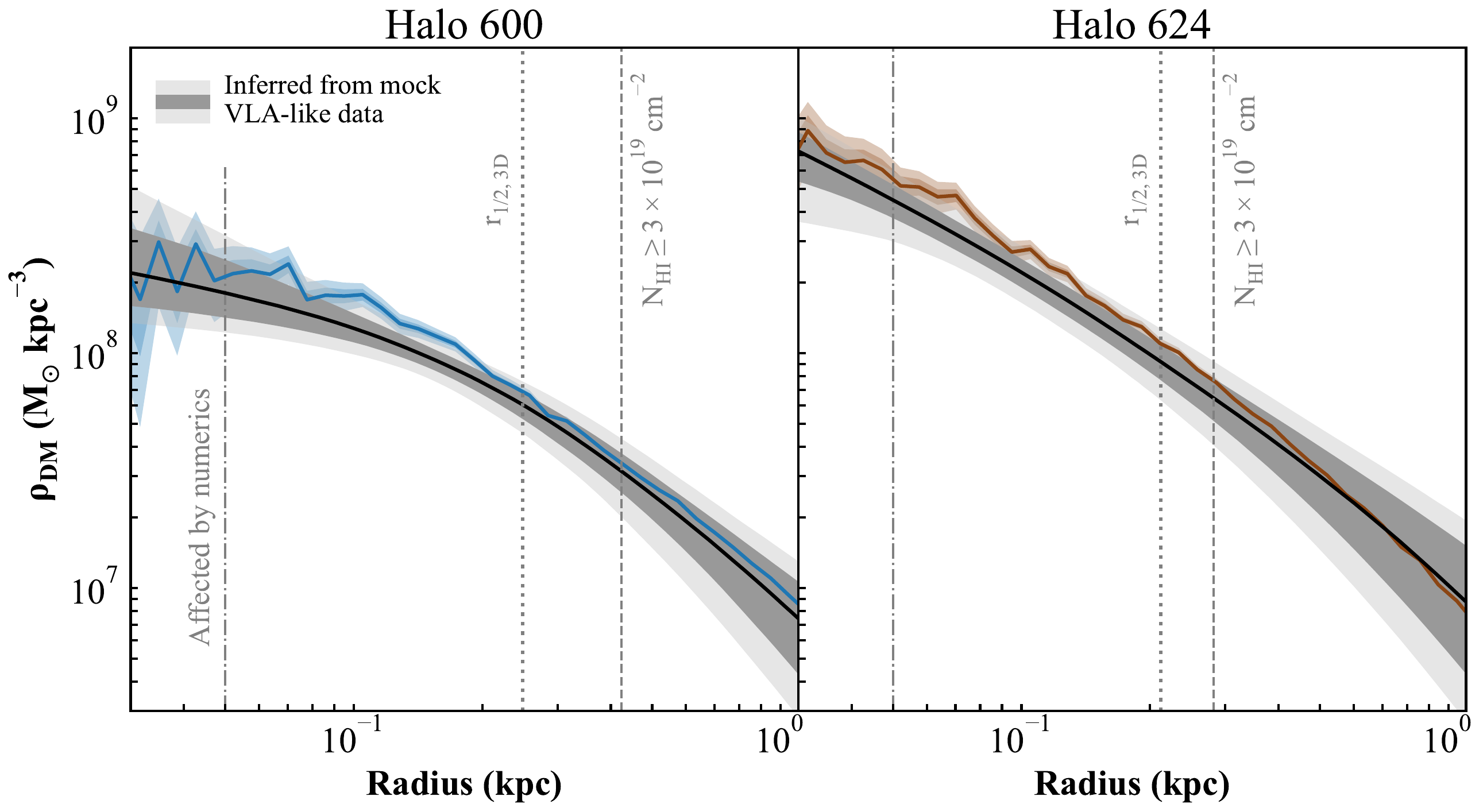}

    \caption{Inferred dark matter density profiles (black line showing the median, grey contours the 1 and 2$\sigma$ intervals) from the rotation curves in Figure~\ref{fig:shortliveddisc} and Figure~\ref{fig:rotationcurve} (left and right, respectively). Both dark matter profiles are recovered within the 95\% confidence intervals, although the inferred dark matter density profile of the star-forming dwarf (left) is slightly flatter than its true profile (blue), with all other parameters (e.g. mass, concentration) poorly constrained (Figure~\ref{fig:inferencemarginals}). In contrast, the dark matter density profile is well recovered all the way to the centre in the quiescent galaxy (right), ruling out a dark matter core of size $< 100\, \pc$ at 95\% confidence (Figure~\ref{fig:inferencemarginals}). This highlights the clear prospects offered by quiescent systems, as they are more favourable to host long-lived and organized rotation, that is easier to interpret for dark matter science.
    }
    \label{fig:dminference}

\end{figure*}

Contrastingly, two of our low-mass dwarfs undergo significantly quieter evolution, with several billion years without forming new stars (see also \citealt{Rey2020}). The lack of star-formation activity since $z\approx4$ leads to more stable \hi reservoirs in these systems with better organized kinematics (Figure~\ref{fig:hivelmaps} and~\ref{fig:vphitimeevolution}). In particular, one of our quiescent \hinospace-bearing dwarf showcases a long-lived, close-to-circular \hi rotation curve (Figure~\ref{fig:rotationcurve} and Appendix~\ref{app:longliveddisc}) that could be readily and robustly interpreted for dark matter inferences. We tie the existence of this long-lived rotation curve to the specifically oblate shape of its host dark matter halo, which plays a key role in building the final \hi disc by torquing circumgalactic gas to align with its axisymmetric revolution axis (Figure~\ref{fig:haloshape} and~\ref{fig:angmomorientation}).

Our results point to \hi rotation being generally rare, sensitive and potentially challenging to interpret in faint \hinospace-bearing dwarfs. But we stress that the mere existence of several examples of ordered and easy-to-interpret \hi rotation curves across a suite of only five simulated galaxies is highly promising and strongly motivates further observational and theoretical investigations. In particular, our findings highlight clear avenues to find `golden eggs' enabling robust dark matter inferences, that is  targeting low-mass dwarfs that (i) have been quiescent for an extended period of time and have avoided rapid disruption of their gas flows by stellar feedback from newborn stars; and (ii) are hosted in an oblate dark matter halo whose axisymmetric geometry promotes disc formation.

An extended gap in star formation and a quiescent period can be inferred from a color-magnitude diagram and a lack of young, blue stars when deep photometric imaging is available. Candidates for such quiescent low-mass dwarf galaxies have in fact already been reported (\citealt{Janesh2019}, although see also \citealt{Rhode2023}) but, unfortunately, the shape of their host dark matter halo cannot be known a priori (or at all). Oblate dark matter halo shapes are statistically rarer amongst the population of high-mass dark matter haloes (e.g. \citealt{Jing2002, Maccio2007, Schneider2012, Bonamigo2015}). But their fraction is steadily rising towards lower halo masses (e.g. $\approx$ 20 per cent of haloes with $\Mvir = 10^{12}\, \Msol$ compared to $\approx$ 10 for $\Mvir = 10^{13}\, \Msol$; \citealt{Vega-Ferrero2017}). Statistical quantifications of halo shapes across the low-mass dwarf galaxy population remain limited in term of sample sizes (e.g. \citealt{Downing2023,Orkney2023}), but these estimates are in line with the (very) small number statistics of one-out-five oblate halo in our suite. Our established link between halo shapes and gas rotation in small dark matter haloes makes quantitatively refining these numbers particularly pressing. 

When detected, observationally characterizing and interpreting \hi rotation curves in such small and faint systems is likely to pose a difficult, but achievable, challenge. We highlight this point in Figure~\ref{fig:dminference}. To this end, we take the pressure-corrected rotation curves presented in Figure~\ref{fig:shortliveddisc} and~\ref{fig:rotationcurve} and resample them with 100 pc spatial resolution ($\approx 10''$ at 2 Mpc, a spatial resolution accessible to interferometric studies in nearby low-mass dwarfs with current instruments; \citealt{Bernstein-Cooper2014, Adams2018}; M. Jones et al. in preparation). Motivated by the spectral and imaging sensitivity of these same studies, we then assume Gaussian velocity errors of $0.8 \, \kmpers$ and that the rotation curve is imaged down to a column density of $\NhiaboveXX{5}{19}$. We then fit these mock rotation curves with `coreNFW' dark matter density profiles using the Bayesian approach described in \citet{Read2016DwarfRCs} (see also \citealt{Read2017, Read2019DMHeating}). Appendix~\ref{app:inference} describes this procedure in more detail and shows the posterior dark matter halo masses, concentrations and core sizes for each dwarf galaxy. Figure~\ref{fig:dminference} presents the inferred dark matter density profiles, with their median and 1-2$\sigma$ intervals (black line and grey contours) compared to the `true' simulated profiles (coloured contours, with shading showing 1-2$\sigma$ uncertainties from the Poisson noise in each radial bin). 

For both the star-forming galaxy (left) and the quiescent galaxy (right), the inferred dark matter profiles are compatible with their underlying simulated `truth' at $2\sigma$. In both cases, this is mainly driven by an underestimated normalization of the density profile (i.e. the overall dark matter halo mass). This entirely reflects the limited radial range of realistic rotation curves, as current studies in this regime could only image the rising part of the rotation curve, leading to largely unconstrained masses and concentrations (Appendix~\ref{app:inference}). We verified that halo masses are much more accurately and precisely recovered if providing the rotation curve down to $\NhiaboveXX{5}{17}$. 

In the star-forming case, the inferred profile is flatter than the truth at $r \leq \rhalflight$, with the inference showing marginal preference for a non-zero core of size $\approx 0.5 \, \rhalflight$ (Figure~\ref{fig:inferencemarginals}). This reflects the challenges of modelling these systems, as feedback-driven non-circular motions or out-of-equilibrium bulk flows (such as those visible in Figure~\ref{fig:shortliveddisc} that lead to the dip in $\vrothi$ at $\approx 200\, \pc$) can bias rotation curves inferences towards flatter profiles (see e.g. discussion in \citealt{Oman2019}). These effects are further compounded by the limited radial range of the rotation curve, and thus their constraining power on other halo parameters.

Nonetheless, in the case of the quiescent galaxy with a well-ordered rotation curve, the inference confidently recovers a steep dark matter profile all the way to small radii, ruling out a dark matter core of size $< 100\, \pc$ at 95\% confidence (Figure~\ref{fig:inferencemarginals}). If confirmed observationally, such steep profile would establish that SN-driven dark matter heating cease to be efficient in small galaxies. This, in turn, would put a strong and clean bound on dark matter models that flatten density profiles in dwarf galaxies (e.g. self-interactions, axions).

This proof-of-concept inference is far from a complete end-to-end validation and makes significant simplifying assumptions (e.g. pressure corrections are derived from simulated profiles rather than a \hi datacube, galaxy inclination is assumed to be perfectly known; see \citealt{Read2017,Oman2019, McQuinn2021,Downing2023} for further discussion of these challenges). Nonetheless, it demonstrates the incredible potential of targeting low-mass galaxies that have been quiescent for a while. These systems not only are more favourable to host long-lived and organized \hi rotation, but their rotation curves are also easier to interpret as star formation and supernovae have had less opportunity to dynamically heat dark matter. Finding these `golden egg' systems will thus be key to unlock high-quality, precise and unbiased dark matter inferences. 

Excitingly, ongoing wide-sky \hi surveys such as Wallaby on the Australian Square Kilometer Array Pathfinder (\citealt{Koribalski2020}), Apertif-Medium deep on the Westerbrook Synthesis Radio Telescope (\citealt{vanCappellen2022}) and efforts with the Five-hundred-meter Aperture Spherical radio Telescope (\citealt{Kang2022}) will soon provide many new candidates for \hinospace-rich, low-mass dwarfs. At the same time, serendipitous detections of such objects are becoming more common (e.g. \citealt{Brunker2019, Hargis2020, Bennet2022,Jones2023PAVODwarf}), particularly in the realm of sensitive \hi surveys cross-matched with deep imaging like MHONGOOSE and MIGHTEE on MeerKat (\citealt{Blok2020, Maddox2021}). These new catalogues and detections will lack the necessary combination of angular and spectral resolution to obtain robust \hi kinematics, but they will be invaluable to select the most promising targets for high-quality rotation curve follow-ups with deep interferometry from VLA and MeerKAT. Combined, these new capabilities will ensure we meet the clear promises offered by the very faint end of the \hinospace-bearing population to constrain dark matter physics.

A key aspect to achieving this goal will be to obtain robust predictions of the connection between dark matter and \hi properties in low-mass dwarf galaxies. This requires us to pinpoint the coupling between a dwarf's ISM and stellar feedback, which is key not only to regulate the ability of supernova-driven outflows to drive dark matter heating, but also their ability to disrupt \hi discs. Many galaxy formation models, including our own, now converge in predicting that the low stellar masses of faint dwarfs does not provide enough SN energy to fully heat their central dark matter into a large and flat dark matter core (e.g. \citealt{Penarrubia2012, DiCintio2014, Chan2015, Onorbe2015, Tollet2016, Lazar2020, Orkney2021}). This study provides the first link between these results and the efficiency of \hi disc formation at this galactic mass-scale. Despite these achievements and the accurate modelling of supernova explosions in our simulations, further quantifications are required to better understand the robustness of our predicted \hi kinematics. In particular, photo-ionization feedback can lead to a more gentle and less explosive regulation of star formation (e.g. \citealt{Agertz2020EDGE,Smith2020PhotoRT}) and could further promote \hi disc formation. Resimulating all of our dwarfs accounting for radiative effects and improved tracking of gas flows over time (\citealt{Cadiou2019}) will be tackled in future work (Rey et al. in preparation), allowing us to pinpoint how gas spirals into and flows out of these sensitive objects.

\section*{Acknowledgements}
MR would like to thank Betsey Adams, Erwin de Blok, Corentin Cadiou and Filippo Fraternali for insightful discussions during the construction of this work and comments on earlier versions of this manuscript. We would like to thank the anonymous referee for a constructive review that improved the quality of the manuscript. MR is supported by the Beecroft Fellowship funded by Adrian Beecroft. MO acknowledges the UKRI Science and Technology Facilities Council (STFC) for support (grant number ST/R505134/1). OA acknowledges support from the Knut and Alice Wallenberg Foundation, the Swedish Research Council (grant number 2019-04659), the Royal Physiographic Society of Lund and the Swedish National Space Agency (SNSA Dnr2023-00164). AP is supported by the Royal Society. AAP acknowledges support of the STFC consolidated grant numbers [ST/S000488/1] and [ST/W000903/1]. WM thanks the Science and Technology Facilities Council (STFC) Centre for Doctoral Training (CDT) in Data intensive Science at the University of Cambridge (STFC grant number 2742968) for a PhD studentship. This project has received funding from the European Union’s Horizon 2020 research and innovation programme under grant agreement number 818085 GMGalaxies. This work was performed using the DiRAC Data Intensive service at Leicester, operated by the University of Leicester IT Services, which forms part of the STFC DiRAC HPC Facility (www.dirac.ac.uk). The equipment was funded by BEIS capital funding via STFC capital grants ST/K000373/1 and ST/R002363/1 and STFC DiRAC Operations grant ST/R001014/1. DiRAC is part of the National e-Infrastructure. The authors acknowledge the use of the UCL Grace High Performance Computing Facility, the Surrey Eureka supercomputer facility, and their associated support services. This work was partially supported by the UCL Cosmoparticle Initiative.

We thank the developers and maintainers of \textsc{pynbody} (\citealt{Pontzen2013}), \textsc{tangos} (\citealt{Pontzen2018}), \textsc{numpy} (\citealt{vanderWalt2011}), \textsc{scipy} (\citealt{Virtanen2020}), \textsc{jupyter} (\citealt{Ragan-Kelley2014}), \textsc{matplotlib} (\citealt{Hunter2007}), the Astrophysics Data Service and the arXiv preprint repository for providing open-source softwares and services that were used extensively in this work.

The main roles of the authors were, using the CRediT (Contribution Roles Taxonomy) system\footnote{\url{https://authorservices.wiley.com/author-resources/Journal-Authors/open-access/credit.html}}: 

MR: Conceptualization ; Data curation; Formal analysis; Investigation; Writing – original draft. MO: Data Curation; Formal analysis; Writing – review and editing. JR: Conceptualization; Formal analysis; Resources; Writing – review and editing. PD: Conceptualization; Writing – review and editing. OA: Methodology; Software; Writing – review and editing. AP: Writing – review and editing. AAP: Writing – review and editing. SK: Conceptualization. WM: Writing – review and editing.

\section*{Data Availability}
The data underlying this article will be shared on reasonable request to the corresponding author.




\bibliographystyle{mnras}
\bibliography{HIrotation} 



\appendix

\section{Pressure support and derivations}\label{app:pressuresupport}
In this Appendix, we revisit the formal framework behind rotation curve analysis to evaluate the validity of common analysis assumptions in the extreme galactic regime that we are considering.

We start from the radial component of the Euler equation in cylindrical coordinates
\begin{equation}
  \frac{\partial \vR}{\partial t} + \vR \frac{\partial \vR}{\partial R} - \frac{\vphi^2}{R} = - \frac{1}{\rho} \frac{\partial P}{\partial R} - \frac{\partial \Phi}{\partial R} \, ,
\end{equation}
where $\vR$ and $\vphi$ are the gas radial and azimuthal velocities, $P$ is the gas pressure, $\Phi$ is the gravitational potential and $R$ the cylindrical radius.

Assuming that the system is in equilibrium ($\partial \vR / \partial t = 0$), that radial non-circular motions are negligible ($\vR \, \partial \vR/\partial R \approx 0$) and defining $\vcirc^2 = R \, \partial \Phi / \partial R$, we obtain
\begin{equation}
  \vcirc^2 = \vphi^2 - \frac{R}{\rho} \frac{\partial P}{\partial R} \, .
  \label{eq:psupporteq}
\end{equation}
Here, $\vcirc$ is sourced by all gravitating components (dark matter, stars and gas) and can be inferred from $\vphi$ by inverting Equation~\ref{eq:psupporteq}. In the absence of pressure gradients throughout the disc, Equation~\ref{eq:psupporteq}, reduces to Keplerian circular motions $\vcirc^2 = \vphi^2$ and the inference is trivial.

But more generally, the measured rotational velocity $\vphi$ needs to be corrected to account for the additional pressure support. This term is often called an asymmetric drift correction due to its similarity with the asymmetric drift derived from Jeans' equations of collisionless dynamics (see \citealt{Binney2008}, subsection 4.8 and \citealt{Pineda2017} for further discussion of the formally distinct assumptions underlying these two derivations).

Since $P$ is not directly observable, we need further assumptions to observationally estimate pressure support across the disc. Further assuming that the gas is ideal and that its pressure is dominated by isotropic processes (i.e. thermal and kinetic turbulence rather than possibly anisotropic non-thermal processes such as magnetic fields or cosmic rays), we can write $P = \rho \, \vdispeff$ where $\vdispeff^2 = c_s^2 + \vdisp^2$. Here $c_s = \sqrt{P / \rho} $ is the isothermal sound speed (applicable for our regime of fast thermal equilibrium) and $\vdisp = \sqrt{\sigma^2_{x} + \sigma^2_{y} + \sigma^2_{z}}$ is the isotropic 3D gas velocity dispersion estimated in each gas cell from the velocities of its 8-closest AMR neighbours\footnote{We use a spline Kernel as implemented by the \textsc{pynbody} library to weight the velocity dispersion, and verified that using an unweighted velocity dispersion, or using the 64 closest AMR neighbours does not impact our conclusions}. 

Injecting these terms in Equation~\ref{eq:psupporteq} then leads to
\begin{equation}
  v_{c}^2 = v_{\phi}^2 - \frac{R}{\rho} \, \frac{\partial (\rho \sigma^2_{\text{eff}})}{\partial R} \, \equiv \vrot ,
\end{equation}
where $\vrot$ is the effective rotational velocity.

Observations can only measure 2D surface densities, $\Sigma_{g}$, rather than the 3D gas density $\rho$. But assuming that the disc structure decouples in the radial and vertical directions (e.g. exponential in $z$ with a scale-height independent of radius), we can obtain
\begin{equation}
  v_{c}^2 = v_{\phi}^2 - \frac{R}{\Sigma_{g} (R)} \, \frac{\partial (\Sigma_{g} (R) \sigma^2_{\text{eff}} (R))}{\partial R} \, .
  \label{eq:psupportapprox}
\end{equation}

Taking $\Sigma_{g} = \Sigma_{\text{\hi}}$ in Equation~\ref{eq:psupportapprox} recovers the commonly used correction to derive $\vrothi$ accounting for pressure support (e.g. \citealt{Bureau2002, Valenzuela2007, Oh2011, Oh2015, Dalcanton2010, Read2016DwarfRCs, Iorio2017, Pineda2017,Oman2019}), where all terms are observable and can be derived from \hi datacubes and their moment 0, 1, and 2 integrals. Due to its $1 / \Sigma_{\text{\hi}} (R)$ scaling, the pressure contribution is expected to be small at higher densities and inner radii, but become increasingly important towards larger radii. Formally, $ - \partial (\Sigma_{\text{\hi}} \vdispeff^2) / \partial R$ can take any sign depending on the radial profiles at hand. But for typical galactic applications, $\Sigma_{\text{\hi}} (R)$ exponentially declines while $\vdispeff (R)$ is more slowly varying across the cold/warm ISM traced by \hinospace, leading to a positive contribution to $v_{\phi}$ in Equation~\ref{eq:psupportapprox}.

In the analysis of this study, we determine 2D radial profiles in a thin slab at $z=0$ height of $\vphi$, $c_s$, $\vdisp$, $\Sigma_{g}$ and smooth them with a 10-pixel Gaussian filter to avoid noise when estimating gradients numerically. We then apply Equation~\ref{eq:psupportapprox} to determine the pressure-corrected $\vrot$ shown as a red dotted line in Figure~\ref{fig:shortliveddisc} and~\ref{fig:rotationcurve} and across Appendix~\ref{app:pressuresupport}.

As discussed in Section~\ref{sec:physicaldrivers}, this traditional correction is adequate and improves the recovery $\vcirc$ from $\vphi$ within  $\rhalflight$ when the \hi distribution is stable over time and well organized (Figure~\ref{fig:rotationcurve}). When active star formation and associated stellar feedback dominate the dynamics of the gas, however, this is much less clear. Assumptions of steady-state equilibrium and circular motions are likely violated by feedback-driven bulk flows, while $c_s$ profiles rising in the outskirts due to the UV background can lead to both positive and negative pressure corrections. This points to a need for new methods to recover dark matter information from star-forming faint dwarfs which we will develop in forthcoming work. 

\section{Additional \hi rotation curves for star-forming dwarfs}\label{app:shortliveddiscs}

\begin{figure*}
  \centering
    \includegraphics[width=\textwidth]{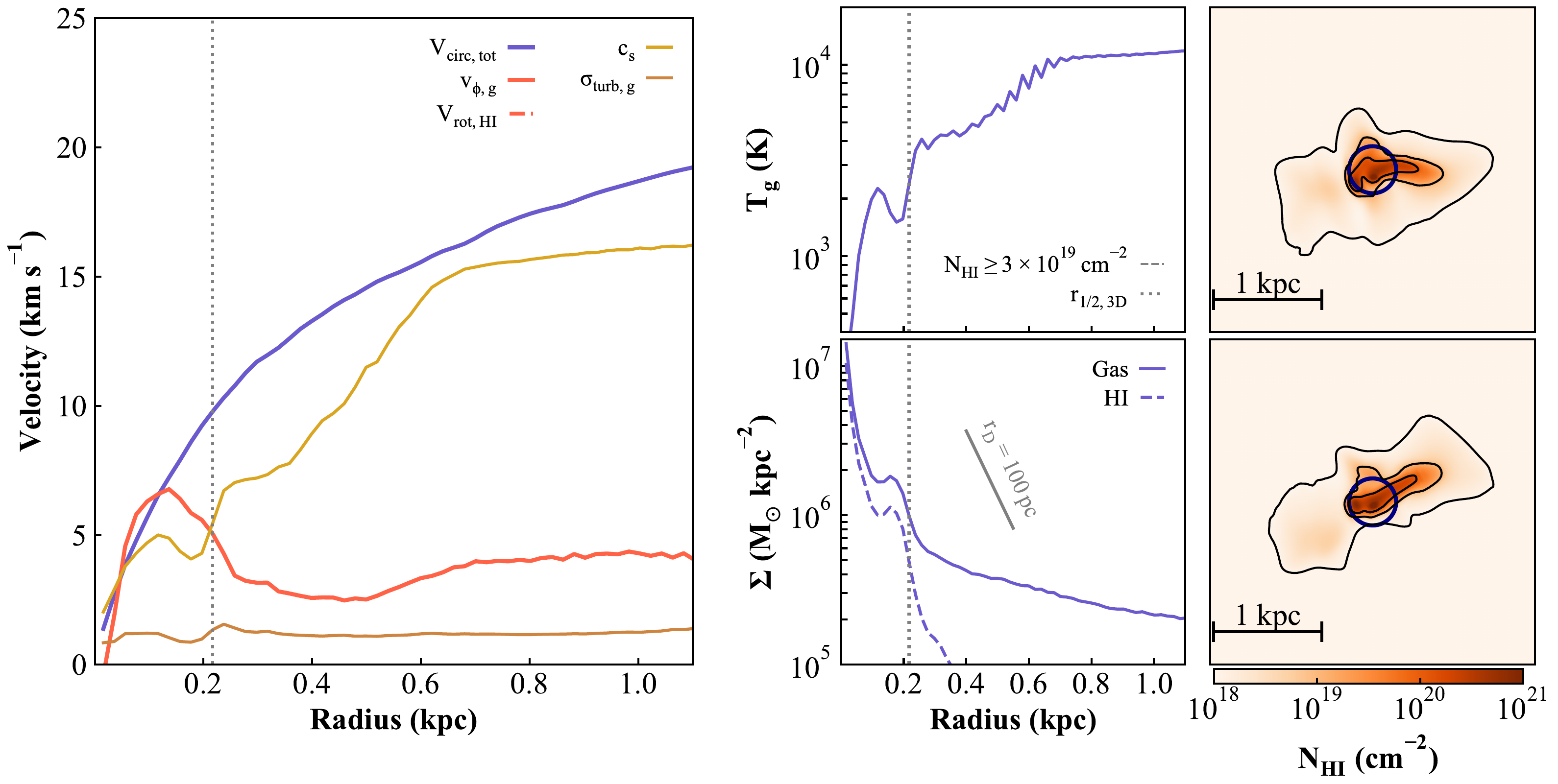}

    \caption{Same object as in Figure~\ref{fig:shortliveddisc} at $t=11.1\, \Gyr$. The \hi distribution is strongly asymmetric, showing a potential small, cold, and rotating \hi disc within $\rhalflight$, but lacking ordered rotation on larger scales.
    }
    \label{fig:halo600cigar}

\end{figure*}

\begin{figure*}
  \centering
    \includegraphics[width=\textwidth]{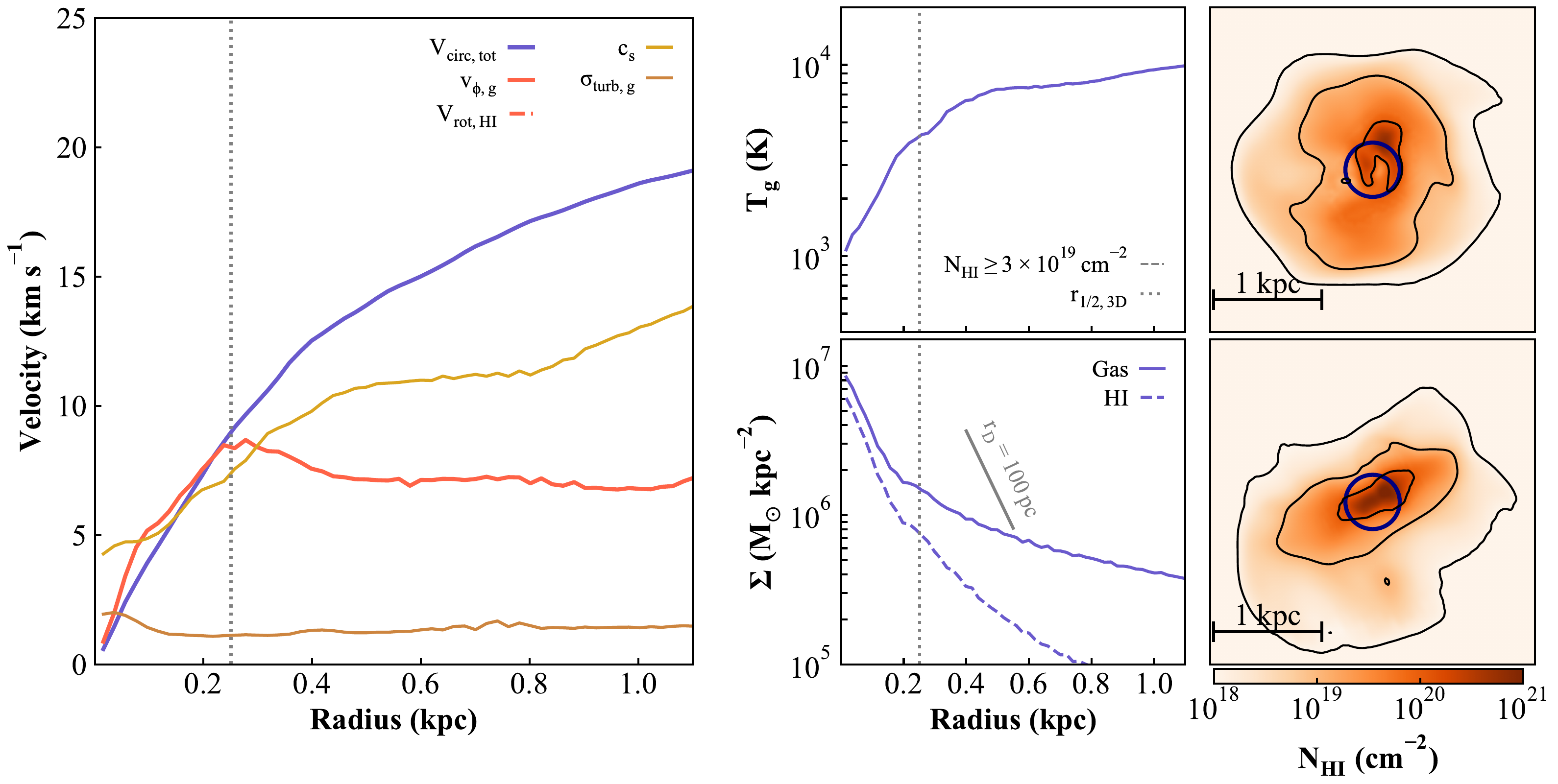}

    \caption{Same object as in Figure~\ref{fig:shortliveddisc} at $t=13.1\, \Gyr$. The \hi distribution is extended and visually flattened on scales larger than $\rhalflight$, with a tangential velocity curve close to the gravitational potential. But the offset between the \hi distribution and the bottom of the potential well (top, right) would likely complicate a dark matter inference.}
    \label{fig:halo600potentialdisc}

\end{figure*}

\begin{figure*}
  \centering
    \includegraphics[width=\textwidth]{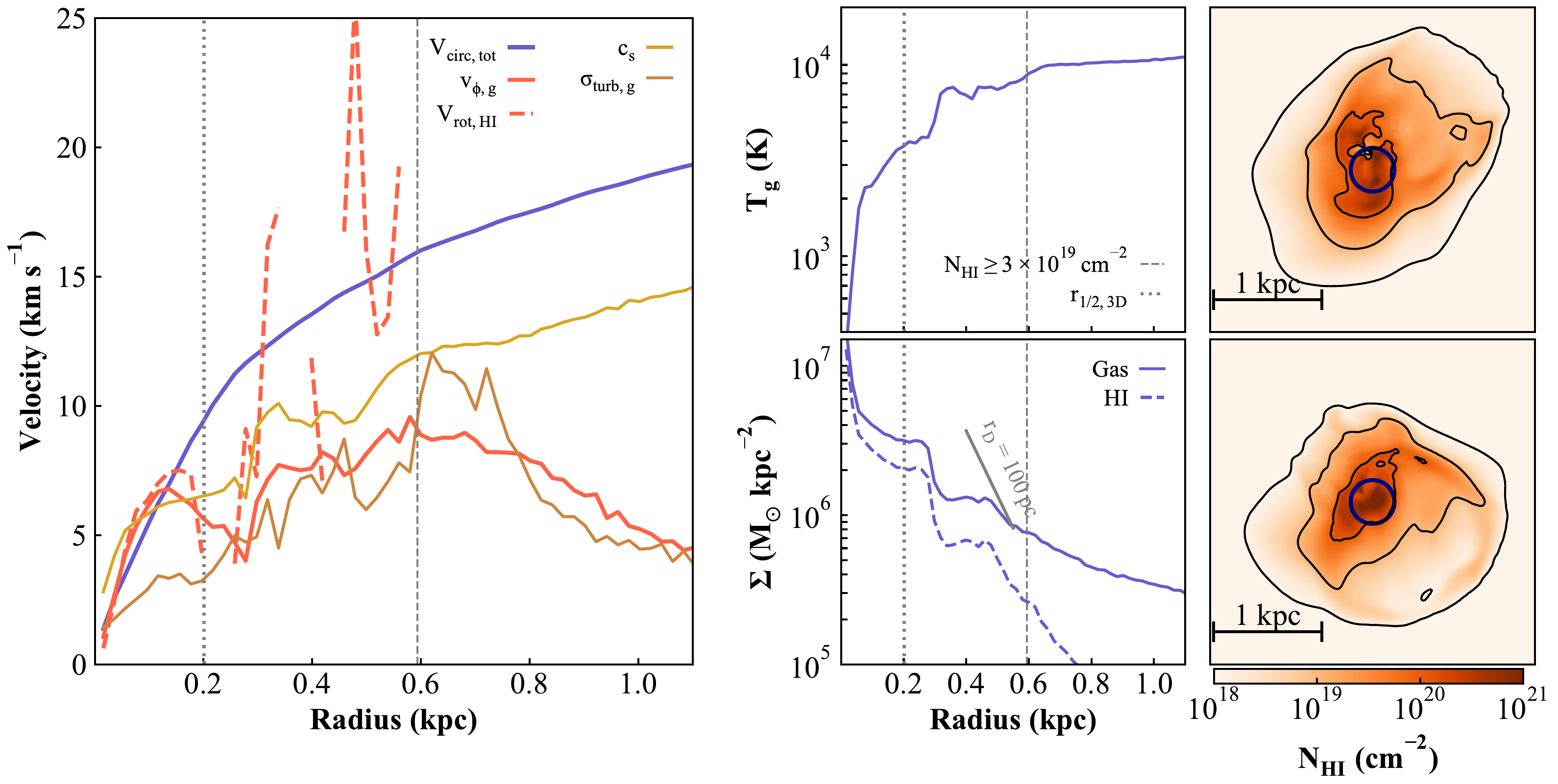}

    \caption{Same object as in Figure~\ref{fig:shortliveddisc} but at $13.2 \, \Gyr$. The \hi distribution is extended (out to $3\, \rhalflight$) but strongly affected by stellar feedback events that drive complex spatial structure and irregular rotation and temperature profiles. 
    }
    \label{fig:halo600disturbed}

\end{figure*}

\begin{figure*}
  \centering
    \includegraphics[width=\textwidth]{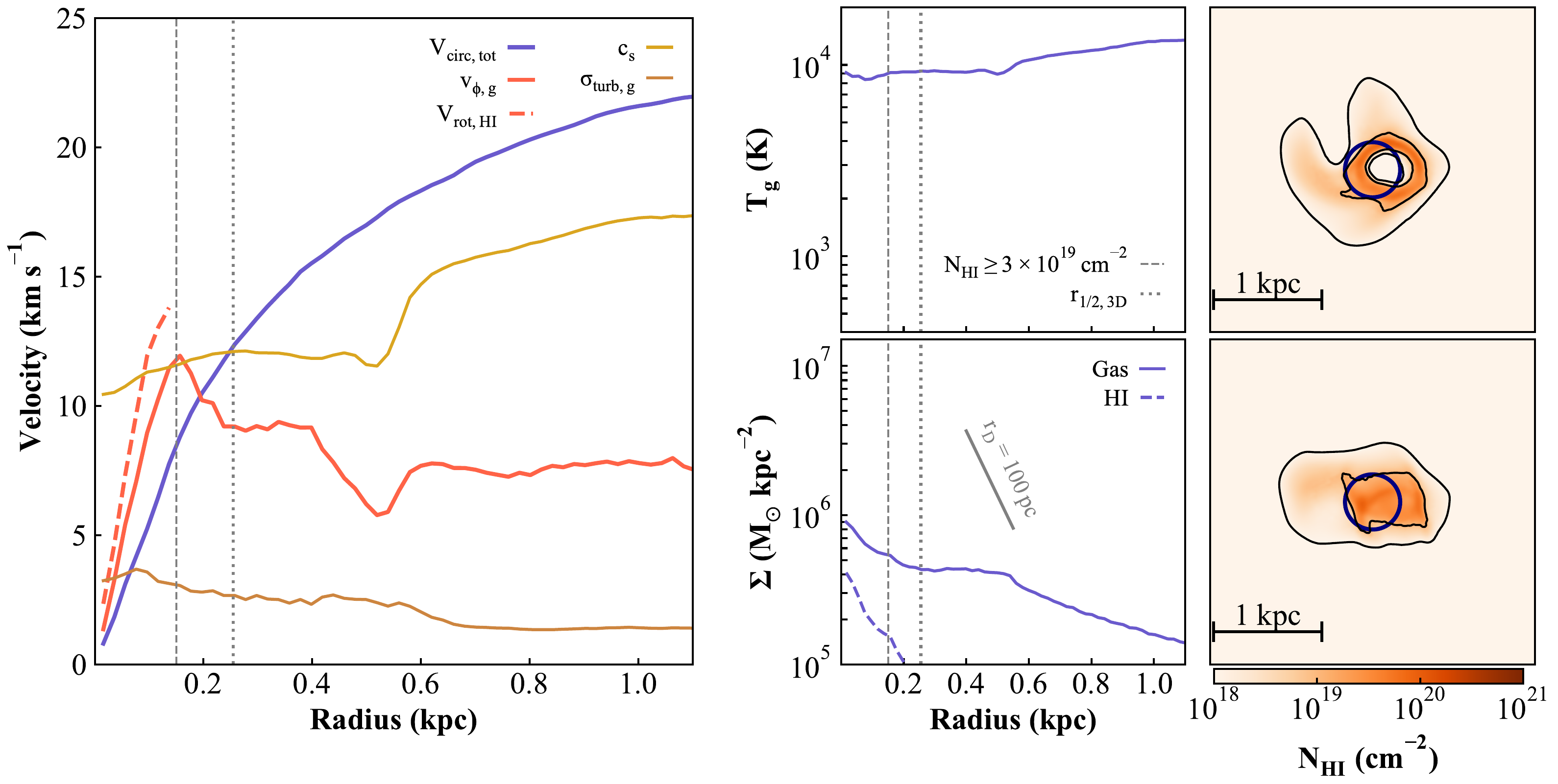}

    \caption{Same as Figure~\ref{fig:shortliveddisc} but showing `Halo 605' at $11.5 \, \Gyr$. The \hi distribution shows signs of rotation (left) that has just been disrupted by a feedback event, driving a large \hi hole in the centre (top, right).}
    \label{fig:halo605hole}

\end{figure*}

\begin{figure*}
  \centering
    \includegraphics[width=\textwidth]{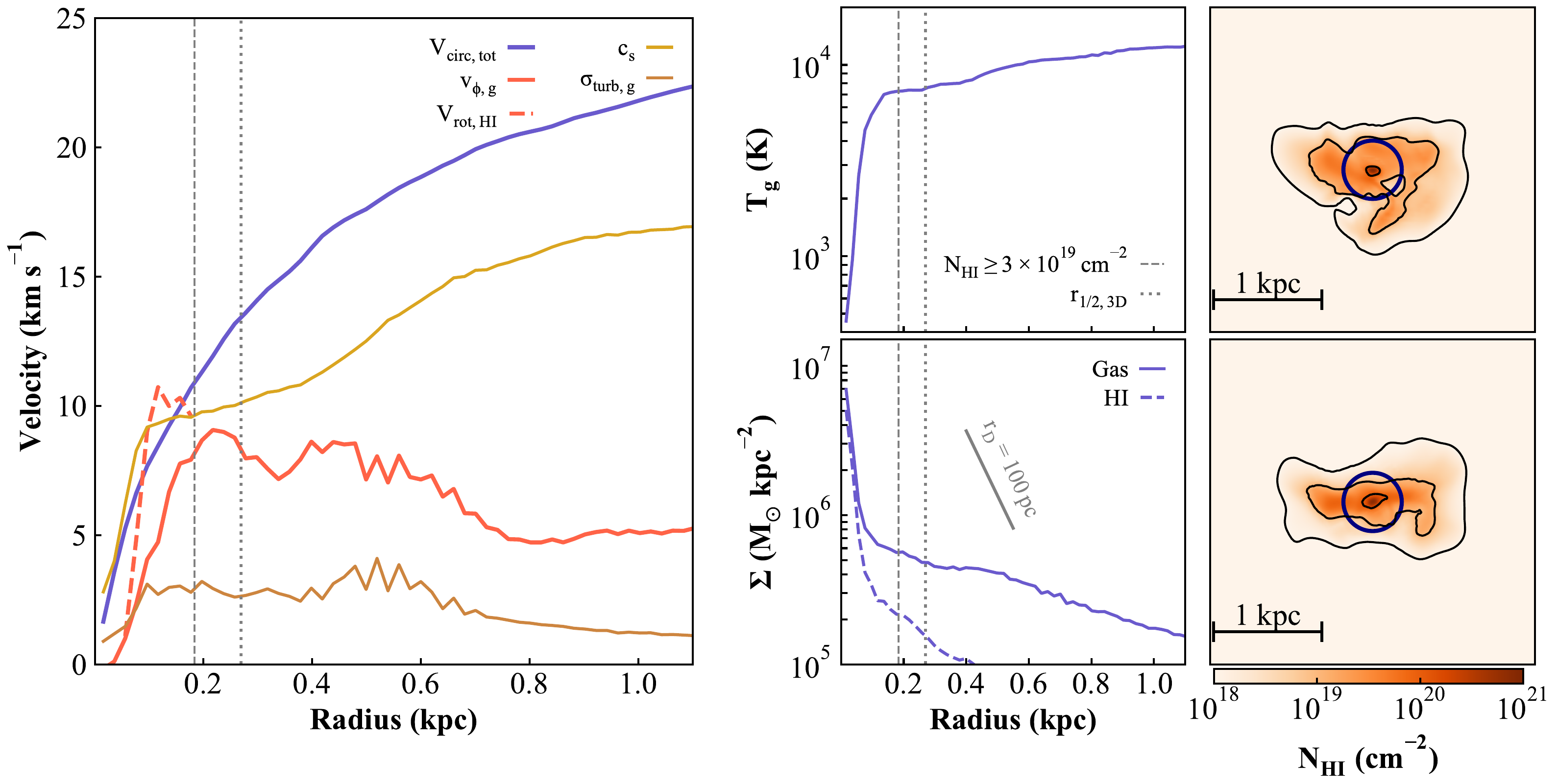}

    \caption{Same object as in Figure~\ref{fig:halo605hole} but at $12.3 \, \Gyr$. The \hi distribution is visually extended and irregular (right panels) with a clearly peaking $\vphi$ profile (left). But this signal is dominated by the thermal support of a warm \hi (top, middle), making the interpretation of $\vrothi$ challenging.}
    \label{fig:halo605disturbeddisc}

\end{figure*}

\begin{figure*}
  \centering
    \includegraphics[width=\textwidth]{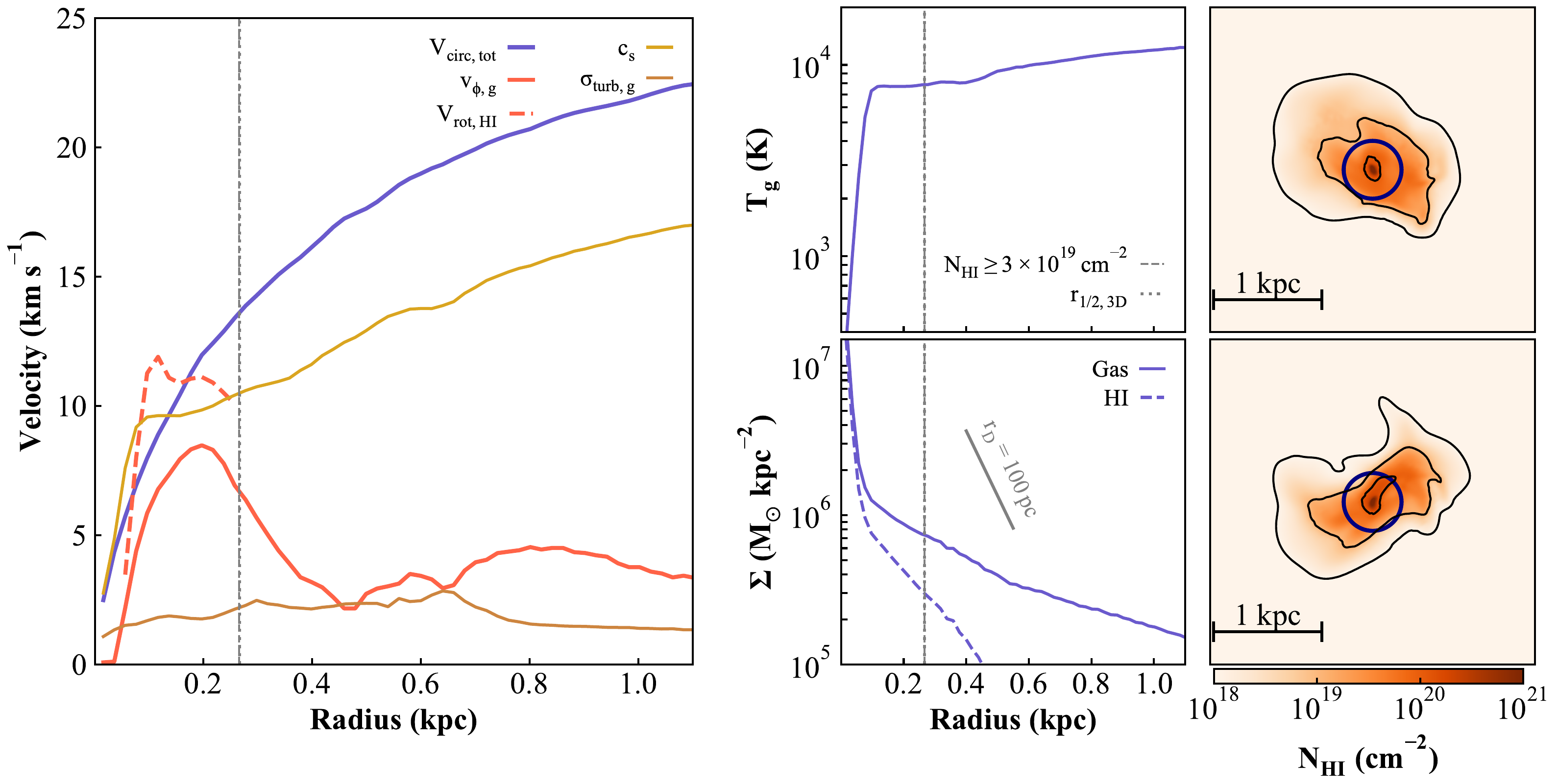}

    \caption{Same object as in Figure~\ref{fig:halo605hole} but at $12.9 \, \Gyr$. Again, the \hi distribution is visually extended and flattened (right panels) but the thermal support of the warm \hi dominates over the rotational signal (left).
    }
    \label{fig:halo605otherdisturbeddisc}

\end{figure*}

\begin{figure*}
  \centering
    \includegraphics[width=\textwidth]{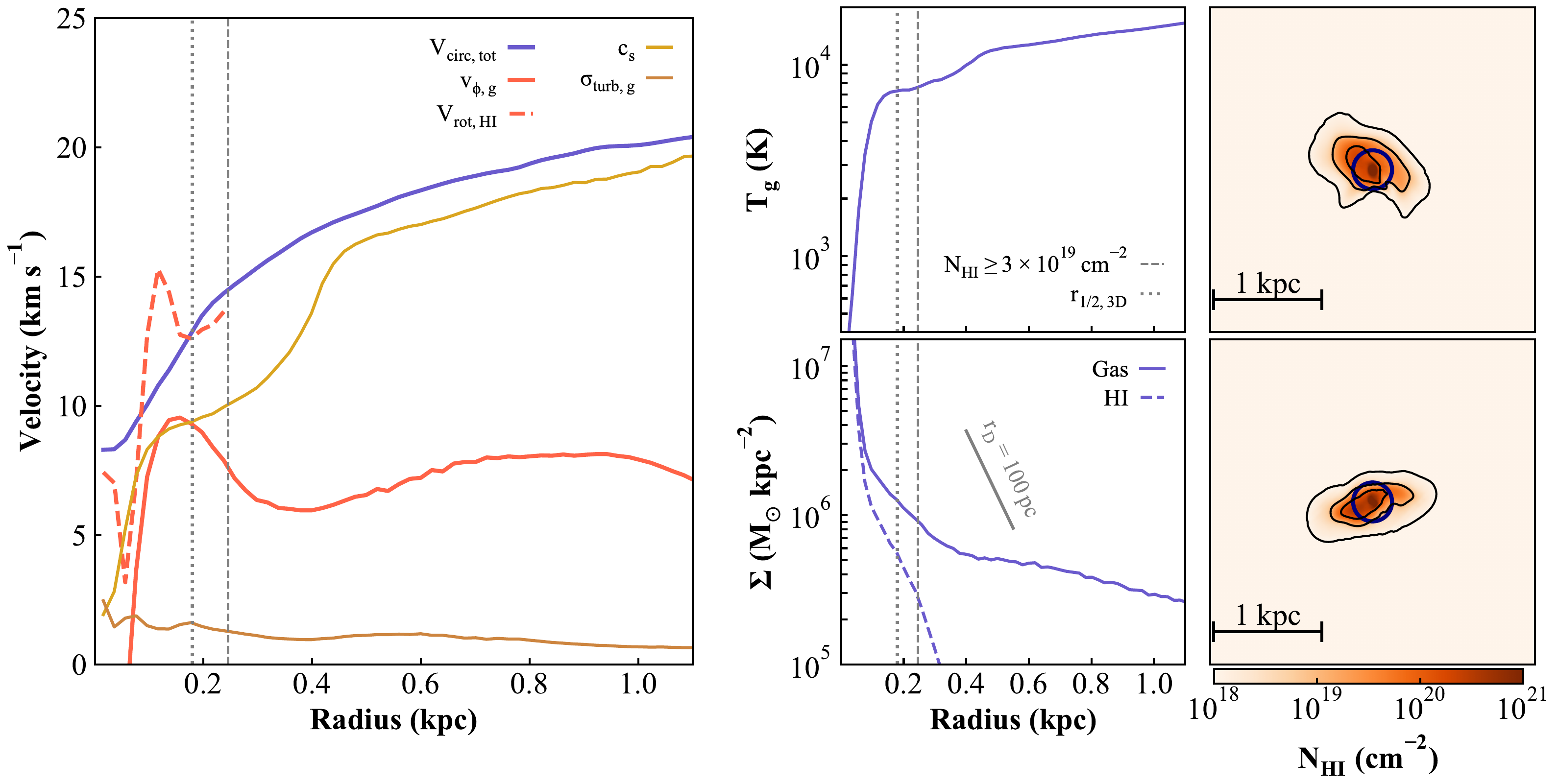}

    \caption{Same as Figure~\ref{fig:shortliveddisc} but showing `Halo 624 GM: Higher final mass' at $t = 5.9 \, \Gyr$. Again, the \hi distribution is visually flattened (right panels) and with a clearly peaking $\vphi$ profile (left). Although the \hi in the centre is cold (top, middle) the interpretation of the rotation curve and $\vrothi$ remains challenging (left).}
    \label{fig:halo624highermasscigar}

\end{figure*}

\begin{figure*}
  \centering
    \includegraphics[width=\textwidth]{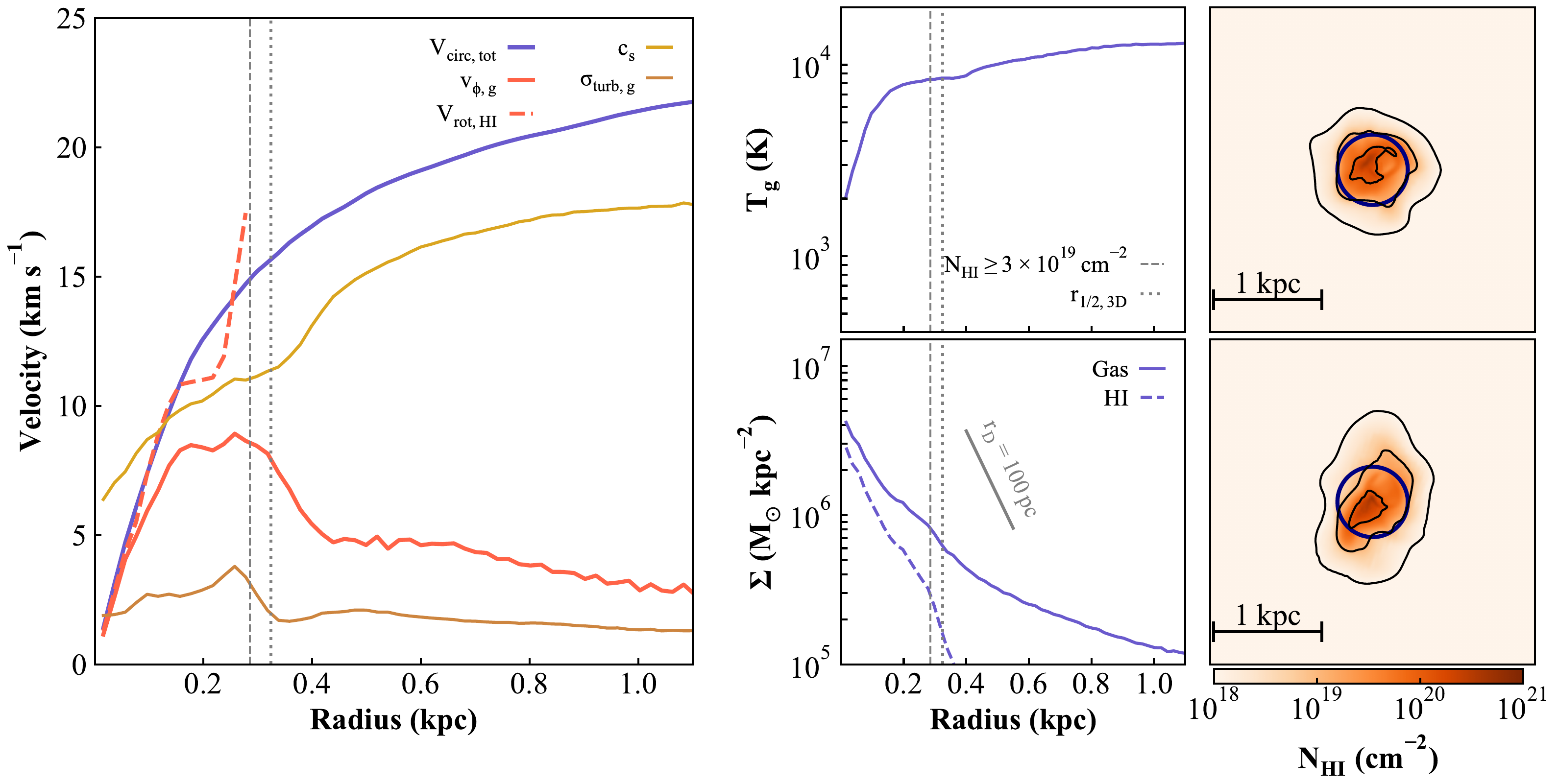}

    \caption{Same as Figure~\ref{fig:halo624highermasscigar} but at $t=13.5 \, \Gyr$. The \hi distribution is fairly regular and comparable in size to the stellar distribution, showcasing an exponential \hi profile. The \hi is warm, with thermal support dominating $\vphi$, but the pressure-corrected $\vrothi$ is an accurate predictor of $\vcirc$ in the very inner part. A feedback event likely just occurred and started dispersing the \hi disc, with its bubble visible in \hi (top, right).
    }
    \label{fig:halo624highermassdisc}

\end{figure*}

In this Appendix, we provide additional visualizations of the rotation curves and \hi column density maps for star-forming dwarfs at the times of higher probability of gas rotation identified in Section~\ref{sec:physicaldrivers} (marked as diamonds in Figure~\ref{fig:vphitimeevolution}). All velocity profiles, rotation curves, radial profiles and \hi column density maps are obtained as described in Section~\ref{sec:kinematics} and match the presentation of Figure~\ref{fig:shortliveddisc}. The following figures show each individual flagged snapshots of star-forming dwarfs and Appendix~\ref{app:longliveddisc} presents further examples of our long-lived rotation curve.

Figures~\ref{fig:halo600cigar}, \ref{fig:halo600potentialdisc} and \ref{fig:halo600disturbed} show three additional timestamps of the same galaxy as in Figure~\ref{fig:shortliveddisc} (`Halo 600'), further demonstrating the time variability and complexity of \hi kinematics driven by efficient stellar feedback in shallow potential wells:
\begin{itemize}
  \item Figure~\ref{fig:halo600cigar} shows a compact and strongly asymmetric \hi morphology, where the dominating rotation signal is difficult to interpret due to the multimodal and offset \hi distribution (see also Figure~\ref{fig:halo624highermasscigar} for another example in a different object).
  \item Figure~\ref{fig:halo600potentialdisc} shows an extended and visually flattened \hi distribution with a $\vphi$ rotation profile (left, red) matching $\vcirc$ (blue) that dominates the velocity dispersion terms (brown and gold). Together with Figures~\ref{fig:shortliveddisc} and~\ref{fig:halo624highermassdisc} in a different object, these are our clearer cases of \hi rotation.
  \item Figure~\ref{fig:halo600disturbed} shows an extended \hi distribution strongly affected by stellar feedback, with high velocity dispersions and $\vrot$ that do not recover $\vcirc$ (see also Figures~\ref{fig:halo605hole}, \ref{fig:halo605disturbeddisc} and \ref{fig:halo605otherdisturbeddisc}).
\end{itemize}

To summarize, our systematic investigation successfully flags short-lived \hi rotation curves in star-forming low-mass dwarfs. Even a single example of an easy-to-interpret rotation curve would prove powerful for dark matter inferences, but we stress that further work is needed to ensure robust inferences. In particular, out-of-equilibrium and non-circular motions driven by stellar feedback seem prevalent in our star-forming low-mass dwarfs and are known to bias inferences of dark matter properties (e.g. \citealt{Read2016DwarfRCs, Oman2019, Downing2023}).

\section{Earlier examples of the long-lived \hi rotation curve}\label{app:longliveddisc}

\begin{figure*}
  \centering
    \includegraphics[width=\textwidth]{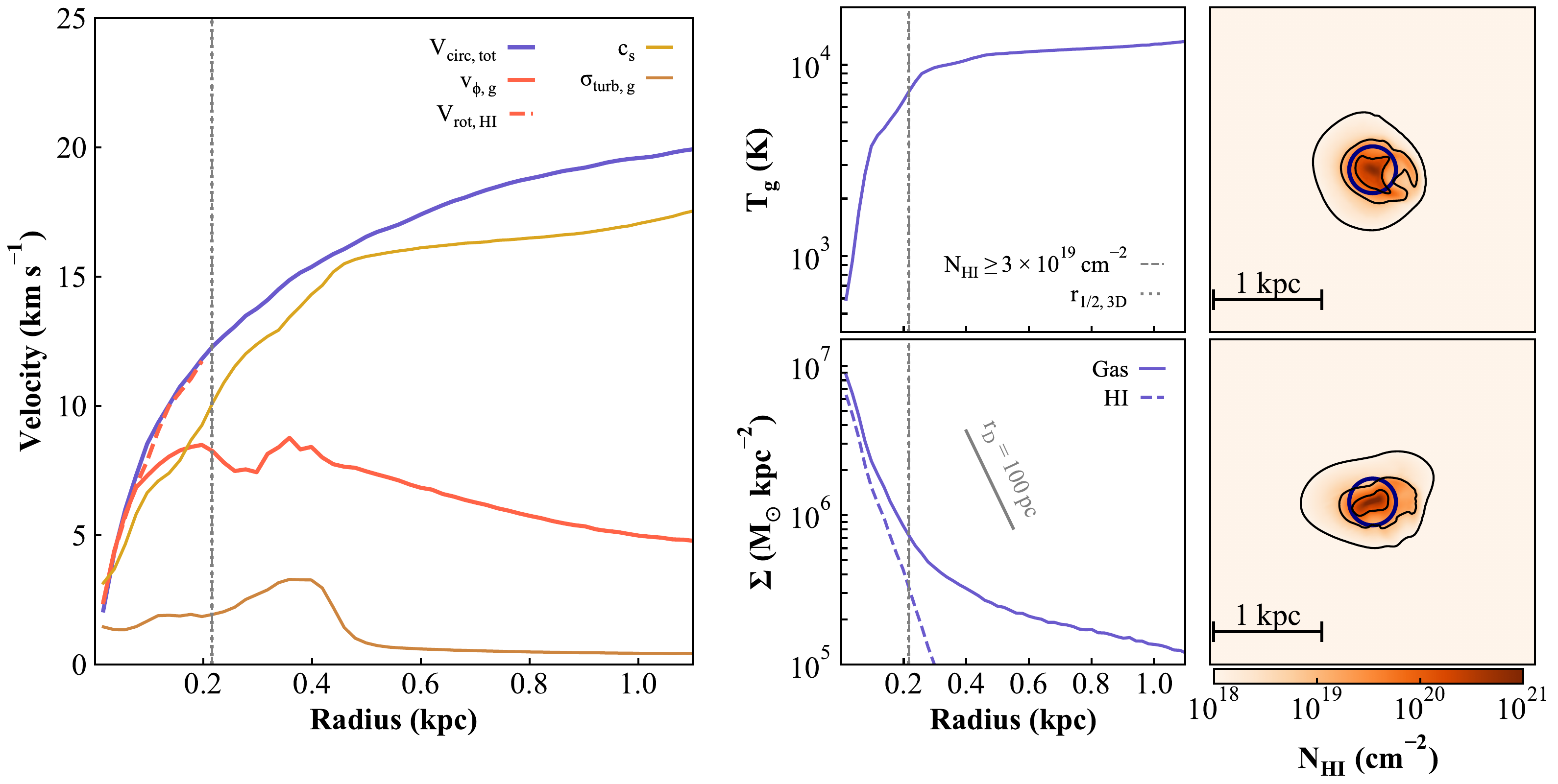}

    \caption{Same as Figure~\ref{fig:rotationcurve} but at $t=11.9 \, \Gyr$. The similarities between the two figures showcase the long-term stability of the clear and easy to interpret rotation curve in `Halo 624'. Feedback from old stellar populations (SNIa, AGB stars; see \citealt{Rey2020}) still inject energy into this small system ($\vcirc \approx 10 \, \kmpers$) and can temporarily disrupt the gas flows (e.g. top, right; see also Figure~\ref{fig:vphitimeevolution}).
    }
    \label{fig:halo624earlierRC}

\end{figure*}

\begin{figure*}
  \centering
    \includegraphics[width=\textwidth]{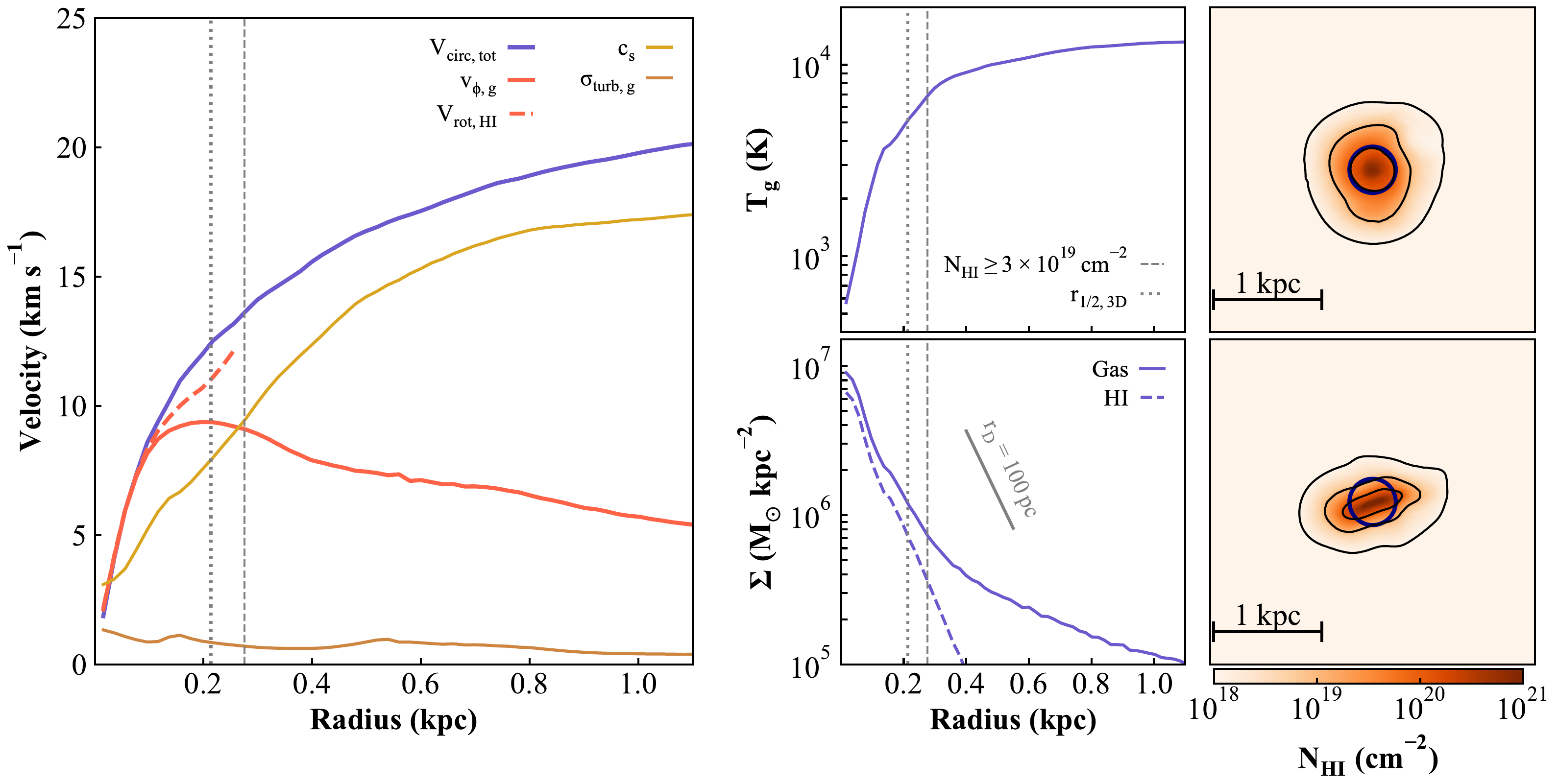}

    \caption{Same as Figure~\ref{fig:rotationcurve} but at $t=13.4 \, \Gyr$, showcasing that the rotation curve observed at $z=0$ has been in place for the last $500\, \Myr$ ($\approx$ 5 full orbits at $300\, \pc$).
    }
    \label{fig:halo624otherearlierRC}

\end{figure*}

In this Appendix, we demonstrate that the rotation curve presented in Figure~\ref{fig:rotationcurve} at $z=0$ is long-lived. Figures~\ref{fig:halo624earlierRC} and~\ref{fig:halo624otherearlierRC} show the velocity profiles, surface density profiles and \hi column density maps at $t=11.9\, \Gyr$ and $13.4 \, \Gyr$. In both cases, we recover the flattened, close-to-exponential \hi distributions with a clear rotation signal that accurately recover $\vcirc$ once pressure support is accounted for ($\vrothi$). 

This rotation curve is thus in place over several billion years, making it a prime target for dark matter science. Gas can still be disrupted by stellar feedback from old stellar populations (e.g. SNIa and AGB stars; see \citealt{Rey2020} for further discussion), with the gap in rotation around $12.0\, \Gyr$ being due to a single SNIa explosion. But these events become intrinsically rarer towards $z=0$ and, as expected from the torques exerted by the host dark matter halo (Section~\ref{sec:sec:shape}), the gas disc rapidly reforms to host stable \hi rotation. 

\section{Dark matter density profile inference} \label{app:inference}

\begin{figure}
  \centering
    \includegraphics[width=\columnwidth]{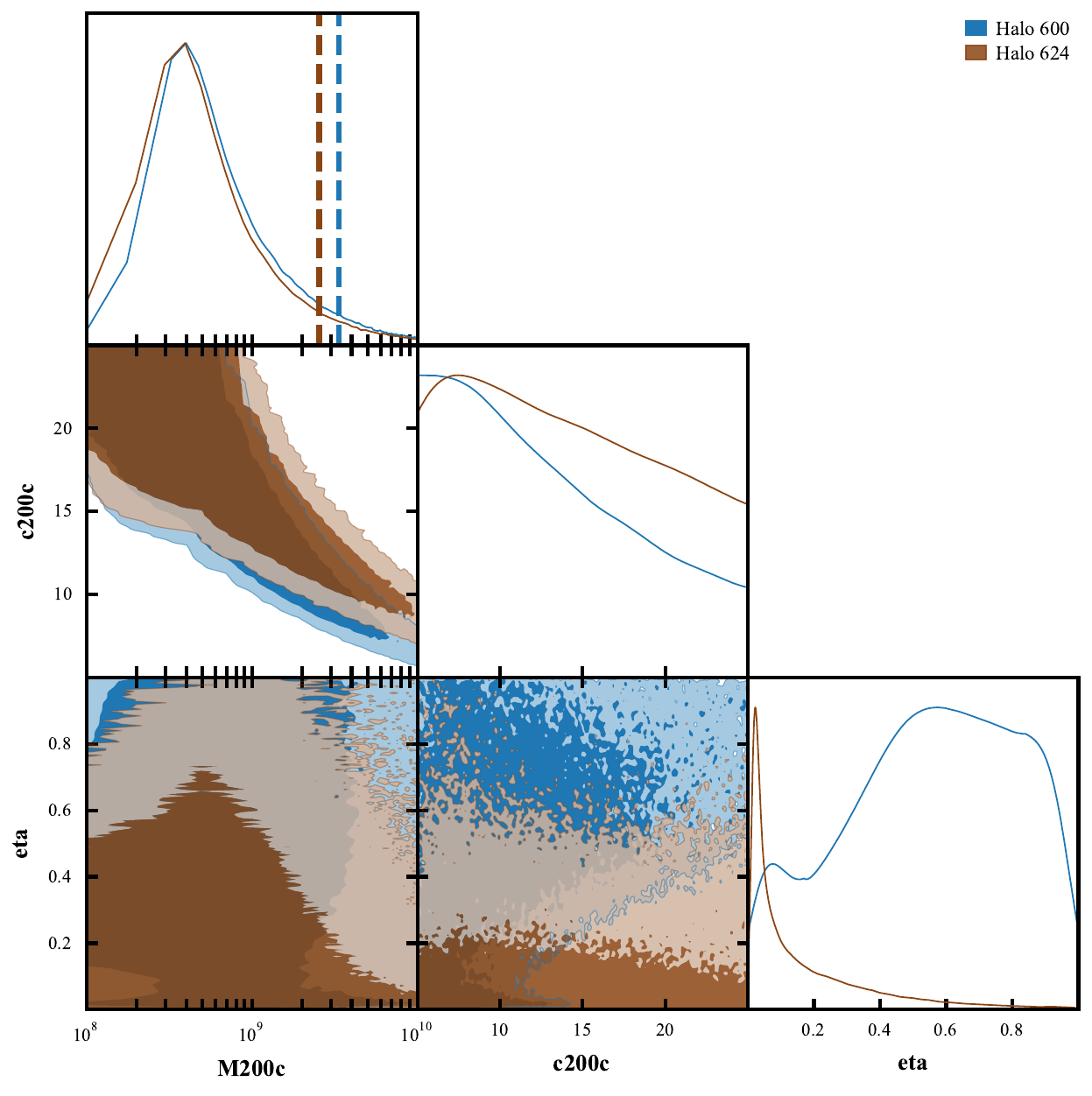}

    \caption{Marginal distributions for the halo mass, halo concentration and core size compared to $\rhalflight$ inferred from the star-forming and quiescent dwarfs (blue and brown contours, respectively). Halo masses and concentrations are poorly constrained as our mock rotation curves only include the inner rise. But this is enough to confidently infer a steep cusp in our quiescent dwarf ($\eta \approx 0$). 
    }
    \label{fig:inferencemarginals}

\end{figure}

In this Appendix, we describe the setup used to infer the dark matter density profiles presented in Figure~\ref{fig:dminference}.

We construct mock data from the rotation curves of Figure~\ref{fig:shortliveddisc} and ~\ref{fig:rotationcurve} corrected for pressure support ($\vrothi$, dashed red). We resample these rotation curves with a spatial sampling of $100 \, \pc$ (i.e. 10'' at $2\, \Mpc$), assign Gaussian errors of $0.8\, \kmpers$ to each velocity point and truncate the rotation curve at $\NhiaboveXX{5}{19}$. All these numbers are motivated by the sensitivity, velocity and spatial resolution currently achievable by deep studies with for example the VLA or MeerKAT (e.g. \citealt{Bernstein-Cooper2014, Adams2018}). 

We fit these mock rotation curve with `coreNFW' profiles using the inference framework of \citet{Read2016DwarfRCs} (see also \citealt{Read2017,Read2019DMHeating} for similar inferences). The `coreNFW' profile has four parameters: the NFW virial mass $\Mvir$ and halo concentration $c_{200c}$, the core parameter $\eta$ that encodes a characteristic size of the dark matter flattening in units of $\rhalflight$, and $\kappa$ encoding the slope of this flattening. All analytics and details about the `coreNFW' profile are described in \citet{Read2016CoresAllTheWay}.   

The Bayesian inference is performed using Markov-Chain Monte Carlo from the \textsc{emcee} package (\citealt{Foreman-Mackey2013}). We run the chain until convergence (50 autocorrelation times) and discard the first 10\% samples as burn in. We assume wide and flat priors on $\Mvir$ and $c_{200c}$ ($10^8 \leq \Mvir \leq 10^{10} \, \Msol$, $5 \leq c_{200c} \leq 30$), and a flat prior of $0 \leq \eta \leq 1$. These priors reflect the large uncertainties in the potential halo masses hosting such  dwarfs (e.g. \citealt{Read2017, Jethwa2018, Nadler2020}), and that in $\Lambda$CDM, such low-$\Mstar$ are not expected to form large cores (e.g. \citealt{DiCintio2014, Tollet2016,Lazar2020,Orkney2021}). To remain consistent with \citet{Read2016DwarfRCs, Read2017,Read2019DMHeating}, we keep $\kappa = 0.04$ fixed during the inference, i.e. assuming that if there is flattening, it is a constant density, flat core. We obtain good fits to the density profiles with this assumption, but we remark that dwarf galaxies in our regime are likely to have profiles shallower than NFW, but not actually flat (i.e. incomplete cusp-core transformation; \citealt{Orkney2021}, Figure~\ref{fig:dmdensity_profiles}). We leave to future work an explorative inference with the $\kappa$ parameter left free, and an investigation of its degeneracies with other parameters.

Figure~\ref{fig:inferencemarginals} shows the result of the inference, with Fig~\ref{fig:dminference} showing the marginalized dark matter density profiles. The mass and concentration are poorly constrained in both cases, although compatible at $1\sigma$ with their true values (marked by vertical dashed lines). This is to be expected as rotation curves extending to $\NhiaboveXX{5}{19}$ are still in the rising phase, making it difficult to break degeneracies between higher-mass haloes with lower concentration and lower-mass haloes with higher concentrations. We checked that this degeneracy is strongly alleviated if providing the full rotation curve ($\vcirc$ in Fig~\ref{fig:rotationcurve}) over an extended radius range ($\NhiaboveXX{5}{17}$). 

The star-forming galaxy has a marginal preference for a core ($\eta = 0.57^{+0.27}_{-0.32}$) that makes the inferred profile lower than its simulated truth at $r\leq \rhalflight$ (Figure~\ref{fig:dminference}). However, the posteriors remain wide and parameters are poorly constrained compared to their input priors. And, if observed, the visibly disturbed \hi distribution of this object (Figure~\ref{fig:shortliveddisc}) would likely make any inference on the absence or presence of cores inconclusive. 

The mass and concentration are similarly poorly constrained for the quiescent dwarf (brown contours), but by contrast, the presence of a steep cusp is confidently preferred ($\eta = 0.08^{+0.35}_{-0.02}$). As we emphasize in Section~\ref{sec:conclusion}, even with current instruments that can only realistically capture the inner rise of the rotation curve in such faint systems, this is enough to recover a steep cusp that would put strong bounds on alternative models of dark matter. 


\bsp	
\label{lastpage}
\end{document}